\documentclass[traditabstract]{aa} 
\usepackage{graphicx}
\usepackage{longtable}
\usepackage{txfonts}
\usepackage{natbib}

\def\suzaku{\textit {Suzaku}}

\def\xmm{\textit {XMM}}
\def\xmmN{\textit {XMM-Newton}}

\def\chandra{\textit {Chandra}}
\def\asca{\textit {ASCA}}

\def\sgr{$\rm Sgr\, A^{\star}$}
\def\sgra{Sgr~A}
\def\sgrb{Sgr~B}
\def\sgrc{Sgr~C}
\def\sgrd{Sgr~D}
\def\fe{Fe K$\alpha$}
\def\Fe{Fe K$\alpha$}
\def\six{6.4~keV}
\def\seven{6.7~keV}

\begin{document}

%
   \title{An X-ray survey of the central molecular zone: \\ variability of the \fe\ emission line
        }

   \author{R. Terrier\inst{1}\thanks{\email{rterrier@apc.in2p3.fr}}
         \and M. Clavel\inst{2} 
         \and S. Soldi\inst{1}
         \and A. Goldwurm\inst{1,3}
         \and G. Ponti\inst{4}
         \and M.~R. Morris\inst{5}
         \and D. Chuard\inst{1,3}
   }

   \institute{APC, Universit\'e Paris Diderot, CNRS/IN2P3, CEA/Irfu, Observatoire de Paris,
     Sorbonne Paris Cit\'e, 10 rue Alice Domon et L\'eonie Duquet, 
     75205 Paris Cedex 13, France
             \and Space Sciences Laboratory, 7 Gauss Way, University of California, Berkeley, CA 94720-7450, USA
             \and Service d'Astrophysique/IRFU/DRF, CEA Saclay, Bat. 709, 91191 Gif-sur-Yvette Cedex, France
             \and Max-Planck-Institut f\"ur extraterrestrische Physik, Giessenbachstrasse 1, D-85748, Garching bei M\"unchen, Germany 
             \and Department of Physics and Astronomy, University of California, Los Angeles, CA 90095-1547, USA
             }

   \date{Received / Accepted }

  \abstract
      {
        There is now abundant evidence that the luminosity of the Galactic super-massive black hole (SMBH) has
        not always been as low as it is nowadays. The observation of varying non-thermal diffuse X-ray emission in molecular
        complexes in the central 300 pc has been interpreted as delayed reflection of a past illumination by bright
        outbursts of the SMBH. The observation of different variability timescales of the reflected emission in the Sgr A
        molecular complex can be well explained if the X-ray emission of at least two distinct and relatively short events
        (i.e. about 10 years or less) is currently propagating through the region. The number of such events
        or the presence of a long-duration illumination are open questions.
        
        Variability of the reflected emission all over of the central 300 pc, in particular in the 6.4~keV Fe K$\alpha$ line,
        can bring strong constraints.
        To do so we performed a deep scan of the inner 300 pc with \xmmN\ in 2012. Together with all the archive data
        taken over the course of the mission, and in particular a similar albeit more shallow scan performed in 2000--2001,
        this allows for a detailed study of variability of the 6.4~keV line emission in the region, which we present here.

        We show that the overall 6.4 keV emission does not strongly vary on average, but variations are very pronounced on smaller
        scales. In particular, most regions showing bright reflection emission in 2000-2001   significantly decrease
        by 2012. We discuss those regions and present newly illuminated features. The absence of bright steady emission
        argues against the presence of an echo from an event of multi-centennial duration and most, if not all,
        of the emission can likely be explained
        by a limited number of relatively short (i.e. up to 10 years) events.
      }


   \maketitle
   

\section{Introduction}
The supermassive black hole (SMBH) at the centre of our Galaxy is accreting matter at a very low rate \citep{Wang13}. 
Its associated source, \sgr, has a bolometric luminosity of only $\sim 10^{36}$ erg/s which is more than eight orders of magnitude smaller than 
the Eddington luminosity of a SMBH of about 4 million solar mass \citep{2016ApJ...830...17B}. It is highly variable in the infrared
and  experiences regular flares in  X-rays, with a measured frequency of 1.1 per day above $10^{34}$ erg/s
\citep{Neilsen13}. The most dramatic changes in luminosity have been observed in X-rays with flux increases
of more than two orders of magnitude compared to quiescent levels \citep{Goldwurm03,Porquet08,Nowak12,Barriere14,Ponti15b}.

There are many phenomena  that might have been caused by intense past periods of activity of \sgr\ \citep[see e.g.][]{Ponti13}. 
An active phase a few million years ago could explain the existence of the huge Fermi bubbles extending 10 kpc
above and below the Galactic Centre (GC) \citep{Su10,Zubovas11,Guo12} as well as the high level of ionization in the Magellanic
stream observed by \cite{BlandHawthorn13}. Similarly, a large region located 2$^\circ$ below \sgr\, filled
with out-of-equilibrium plasma recombining for about 100 kyr was possibly produced by an energetic event
at the GC about 100 kyr ago \citep{Nakashima13}.

There are also numerous lines of evidence that \sgr\ was much brighter in the more recent past. X-ray light echoes,
mostly through the fluorescent Fe K$\alpha$ line \citep{Sunyaev93,Koyama96}, have been observed
in many locations within the 150~pc radius
region surrounding \sgr\ which is usually referred to as the central molecular zone \citep[CMZ,][]{Morris96}.
This gas is optically thick enough to efficiently scatter X-rays emitted by a source in the GC region. 

Variations of the reflected emission have been observed in a number of distinct places in the CMZ: in the \sgra\ complex
\citep{muno07,ponti10,capelli11,capelli12,clavel13,clavel14b} as well as in \sgrb\ \citep{inui09,terrier10,nobukawa11, Zhang15}. 
Besides establishing the nature of the emission as reflection, these rapid variations of the non-thermal X-ray emission from 
molecular clouds provide fundamental information on the temporal behaviour of the illuminating source. 

The detailed light curves of the emission measured on small scales in the \sgra\ complex with \chandra\ have so far revealed
two distinct time constants in the region: a steady variation over 10 years and a rapid rise and decay over a few years. 
\citet{clavel13} have concluded that at least two distinct events are currently propagating through the \sgra\
complex; one has a typical duration of about 10 years while the other is much shorter, perhaps 2 years or less.
Both are intense, with typical luminosities of the order of $10^{39}$ erg s$^{-1}$ or more, and are likely originating in \sgr.   

Understanding these centennial outbursts of \sgr\, requires knowledge of the number and typical
recurrence rate of events currently being reflected by the CMZ.  If only reflections occurring within particular
molecular complexes are considered, it is difficult to see the full picture: are all reflections due to the same events? Was there
a long-term period of sustained activity, or conversely, were there multiple short events \citep[see e.g.][]{ryu13} ?  To answer
these questions, it is necessary to have a global view of variations in the region. 

To achieve this, \xmmN\ performed a deep (640 ks) scan of the CMZ at the end of 2012. A comparison with older data and 
in particular with a shallower survey performed in the early mission in 2000--2001, makes it possible to assess 
the level of variation in the whole region.    

We present in this paper a systematic comparison of these two datasets. Following the approach of \citet{clavel13},
we determine the \six\ line flux in 1$'$ pixels over the entire CMZ at various epochs to test for variability in a
systematic manner. We show that most of the regions that are bright in this line have experienced significant variation
over the 12 years of \xmmN\ observations. 

Examining the variable emission from a number of molecular complexes, we confirm the presence of variations characterised by
two distinct time constants observed in various parts of the Sgr A complex; we report the first signature of a decrease of the Fe K$\alpha$
flux observed  in Sgr C and discuss the characteristic duration of the associated event; we also provide evidence suggesting
an increase of the emission  from the Sgr D complex between 2000 and 2012. We report the sudden illumination of a
large elongated structure covering more than 25 pc in projection. 
Finally, we argue that these findings support the idea that almost all of the \Fe\ emission can be attributed
to relatively short-duration events (i.e. up to 10--20 years) and that no long-term sustained activity
is required to explain the observations. 


\section{\xmmN\ observations and data analysis}\label{section:}

\begin{figure}[!t]
\centering
\includegraphics[width=0.9\linewidth]{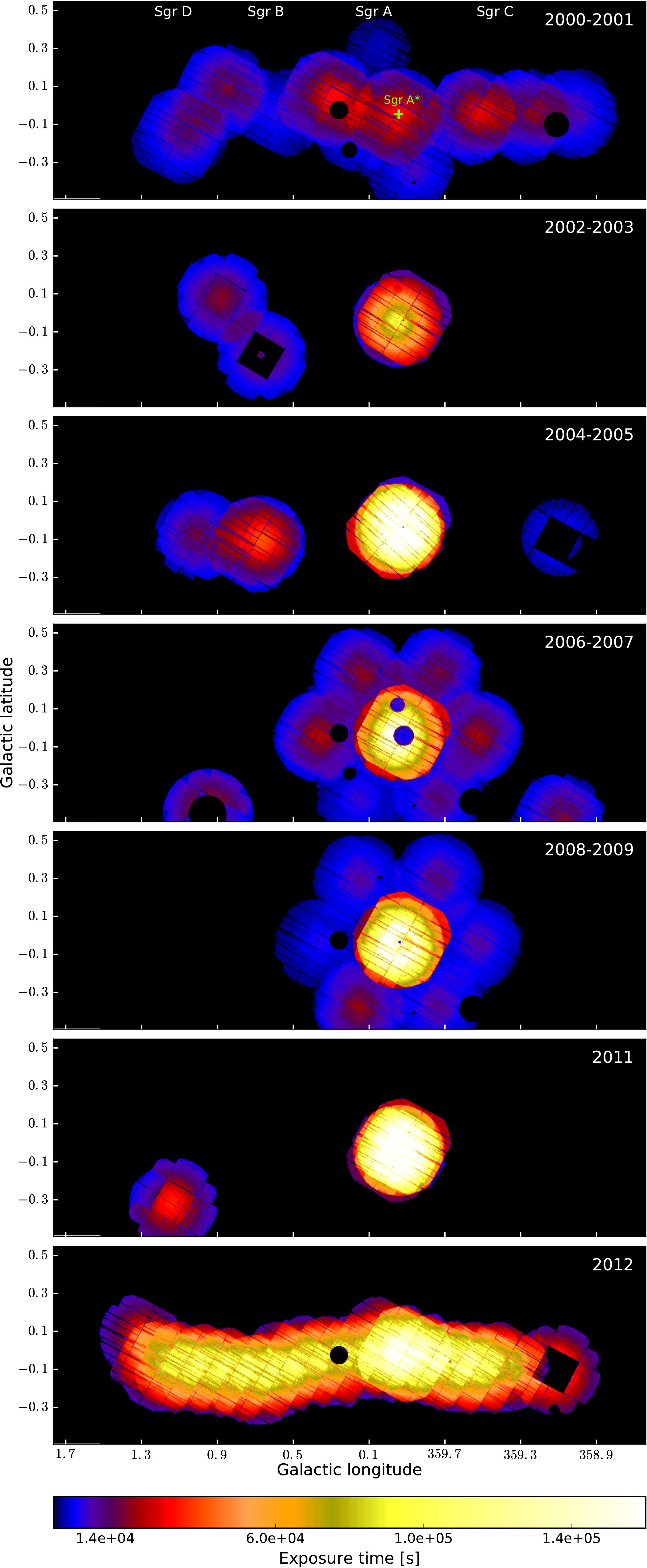}
   \caption{Time exposure maps for the 7 epochs in which the \xmmN\ data have been grouped 
     (in units of seconds, with 2.5$''$ pixel size, and in Galactic coordinates) after applying the exposure time cuts
     discussed in Sect.~\ref{section:analysis}. The darker circular regions correspond to the point sources that have
     been excluded (see the list in Table \ref{tab:trans}). The MOS1 and MOS2 exposures have been rescaled to take into
     account the different MOS and PN effective areas in order to create the PN-equivalent exposure mosaics shown here. 
        }
     \label{Fig:expo}
\end{figure}

Since its launch in 1999, the \xmmN\ satellite has been regularly monitoring the inner GC region, mostly 
focusing on \sgr\ and on pointed observations of bright sources,  but also through two more extended scans covering about 
2$^{\circ}$ in longitude and 0.5$^{\circ}$ in latitude, performed in 2000--2001 and in 2012.
In particular, the 2012 scan uniformly covers the sky between $l = -0.8^{\circ}$ and $l = +1.5^{\circ}$ with 16 pointings of
 40~ks each, evenly separated by 7.8$'$ (bottom panel in Fig.~\ref{Fig:expo}).

 These two scans as well as a large number of observations pointed within 1$^\circ$  were used to produce complete maps of
 the X-ray emission in the CMZ \citep{Ponti15a}. We use here the same dataset which yields 101 observations with data from at
least one EPIC detector, and a total of 285 single-exposure images. 


\subsection{Imaging analysis}\label{section:analysis}

\begin{figure}[!t]
\centering
\includegraphics[width=7.0cm]{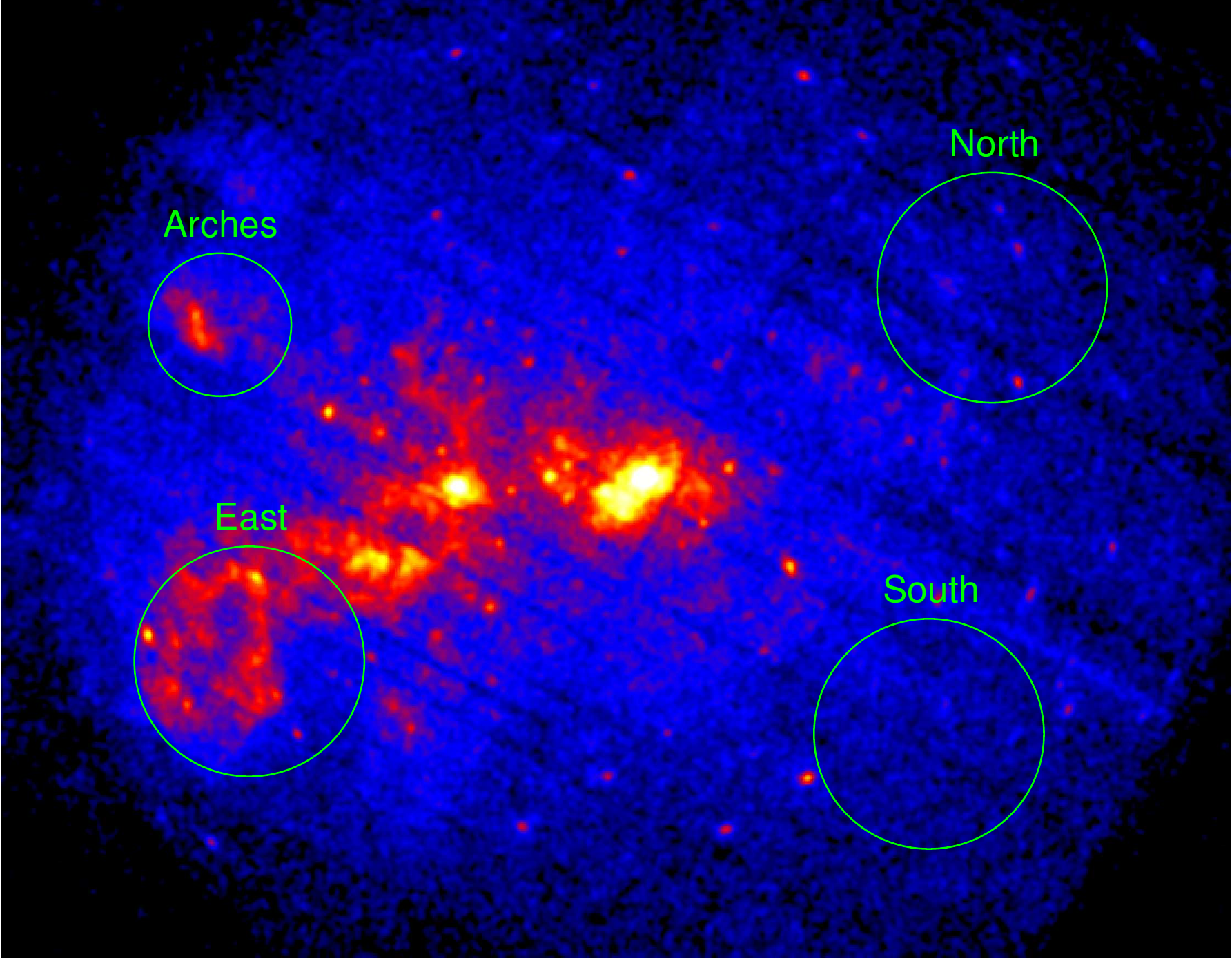}
\includegraphics[width=\linewidth]{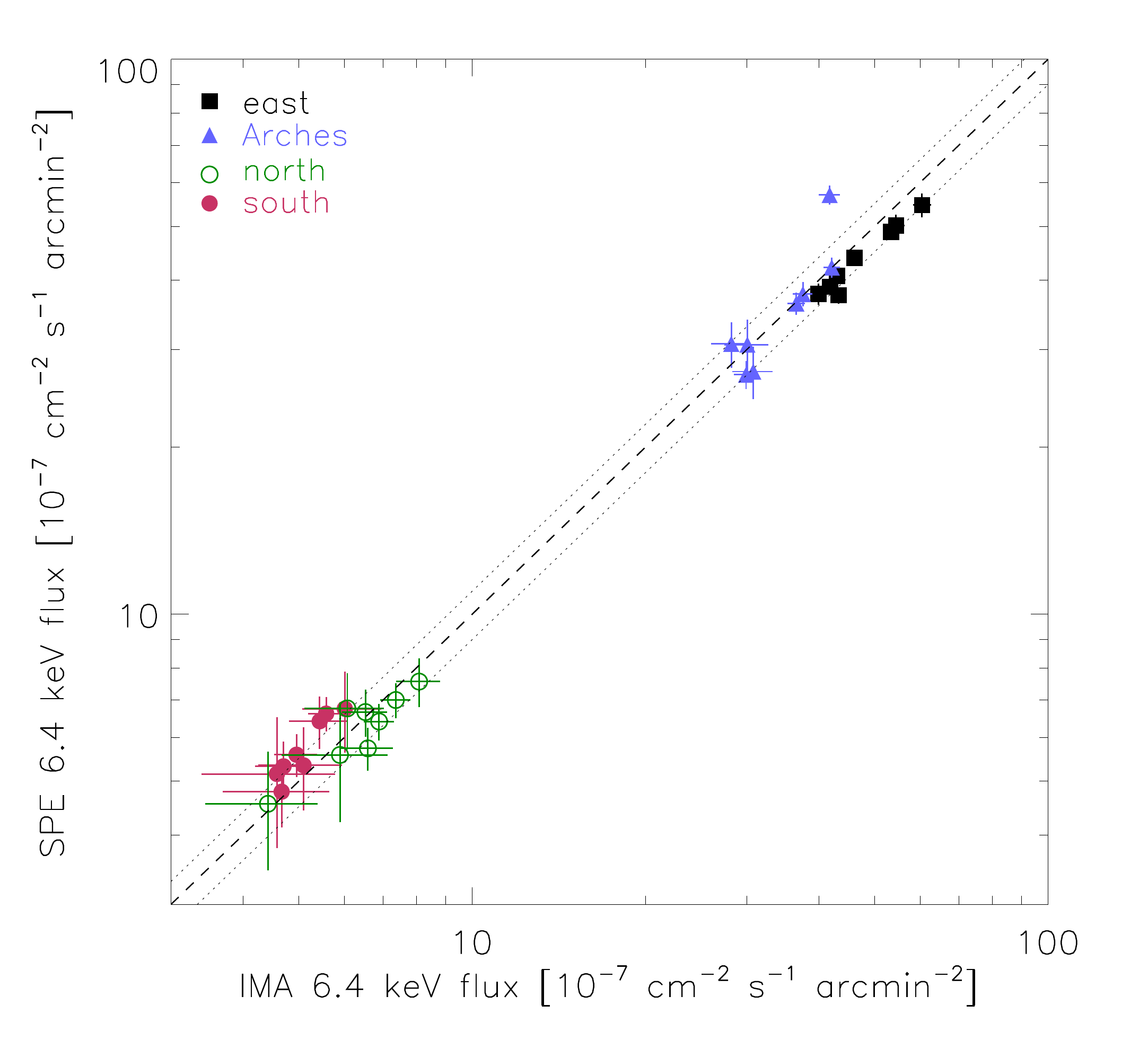}

   \caption{Comparison of the \fe\ fluxes extracted from the mosaic images with those obtained from the spectral 
   fitting for the four test regions within 12$'$ of \sgr visible in the inset. The data have been grouped into eight periods as described 
   in the text (with the 2012 data split between the March and autumn observations). 
   The dashed line indicates where regions with the same fluxes from imaging and spectral analysis should 
   lie and the dotted lines represent the 10\% dispersion around it.
        }
    \label{Fig:ima_spe}
\end{figure}

The selected dataset was analysed in a uniform way using the \xmmN\ Extended Source Analysis Software 
({\tt ESAS}; \citealt{snowden08}) included in the version 12.0.1 of the \xmmN\ Science Analysis Software ({\tt SAS}).
For each available exposure, calibrated event lists were produced using the {\tt SAS} \textit{emchain} 
and \textit{epchain} scripts and periods affected by soft proton flaring were excluded using
{\tt ESAS} \textit{mos-filter} and \textit{pn-filter}. The quiescent particle background 
was obtained using filter wheel closed event lists provided by the {\tt ESAS} calibration database. 

Since the goal of our study is to follow the \fe\ variability across the CMZ 
through a uniform analysis and within different epochs, we based our investigation on the imaging 
(rather than the spectral) analysis of the \xmmN\ data, which allows us to simultaneously study large 
regions, on different angular scales, and to follow the morphological changes of the emission with time.  
Therefore, for each instrument and observation, count, background and exposure time maps were produced in
 the 4.7--6.3~keV continuum energy band and in two narrow bands centred at \six\ and \seven\ 
 ($\Delta E = 160 \rm \, keV$) in order to produce background- and continuum-subtracted images at \six.
  The continuum band was chosen in order to have sufficient signal statistic in an energy 
range not affected by the presence of emission lines from the GC or  instrumental background
 lines.  The continuum underlying the \six\ emission line was estimated using the flux of the 4.7--6.3 
 and 6.62--6.78 keV energy bands and assuming a simple model to represent the X-ray emission spectrum : a power law
  with photon index $\Gamma = 2$ (modelled with \textit{powerlaw} in {\tt Xspec}) and a plasma component (\textit{apec}) 
  with temperature $kT = 6.5 \, \rm keV$ and solar abundances.
Both components were absorbed by a foreground column density of $N_{\rm H} = 7 \times 10^{22} \, \rm cm^{-2}$ \citep{Ponti15a}. 
This allows allows a (background- and continuum-subtracted) net count map to be produced at 6.32--6.48keV for each instrument
and each observation. In order to obtain the total net counts included in the \six\ line, we applied an
efficiency correction to the 6.32--6.48~keV net count map. This correction is required to correct for leakage of events
  outside the selected band because of energy dispersion. The probability that the energy of a line photon
  is measured in the energy range is obtained from the canned redistribution matrices.\footnote{
  \tt{http://xmm2.esac.esa.int/external/xmm\_sw\_cal/calib/\\epic\_files.shtml}. 
  It is position dependent, especially for the EPIC/pn, due to changing energy resolution when moving from the optical axis
  to the detector edges. Maps of the position-dependent correction factor were computed for each instrument based
  on its redistribution matrix. For each observation and instrument, the corrected net line flux is obtained by dividing
  the net counts maps by the corresponding efficiency map reprojected according to the astrometry.}
  
The same correction was also applied to the 6.62--6.78~keV flux map before using it to estimate the continuum 
contribution at \six. For this band,  the correction is applied only to the fraction of flux that is expected to
be contained in the line following the model and parameters considered here (i.e. 82\% of the total \seven\ flux,
while the rest is continuum emission).

To be able to combine images from the different instruments, the exposure time maps were renormalized taking into 
account the different efficiencies of the three EPIC detectors. In addition, circular regions corresponding to some 
of the brightest transient or persistent X-ray point sources have been excluded (see Fig.~\ref{Fig:expo} and
Table~\ref{tab:trans}). 
%
In particular for the transient sources, the corresponding region was excluded only when the source was found to
be in outburst.

Each net count and exposure time map was then reprojected with the \textit{Chandra} tool 
\textit{reproject\_image\_grid} to a common, specified grid. We produced reprojected maps with different 
spatial resolutions using pixel sizes of 2.5$''$ and 1$'$. To produce the final mosaics, summed count 
and summed exposure time maps were produced and the intensity maps were computed.  In order to exclude the less exposed 
regions, for example, the image borders, which might induce spurious results in the variability analysis, we removed: 
1) in each observation, the pixels with exposure times less than 10\% of the maximum in that specific observation, and 
2) in each mosaic, the pixels with exposure times less than 20\% of the median in that specific mosaic.

The variability analysis presented in the following is mostly based on the final mosaic images at \six\ produced by grouping 
the  data taken within the years 2000--2001, 2002--2003, 2004--2005, 2006--2007, 2008-2009,  
2011\footnote{There are no data from 2010 in our data set.}, and 2012. Two mosaics were created for 2012, one with all data 
collected that year (for comparison with the other epochs) and the other with only the 16 pointings of the survey
 (for comparison only with the 2000--2001 scan). Even though the selected time periods have quite different exposure
 coverage, most of the regions of interest for our study are covered for a minimum of 3-4 epochs (e.g. \sgrb, \sgrc, \sgrd) 
and up to 7 epochs for the $15'$ radius around \sgr\ (Fig.~\ref{Fig:expo}).

\subsection{Comparison of the imaging and spectral fluxes}\label{section:ima_spe}
In order to validate the results obtained from the imaging analysis, we compared the \fe\ fluxes measured 
with our procedure to those obtained with the standard {\tt ESAS} spectral analysis for some specific regions. 
For this purpose, we selected four circular regions of different surface brightness within 12$'$ of \sgr.
In particular, two low-brightness regions with a radius of 164$''$ were chosen at the west of \sgr\ 
(regions labelled `north' and `south'), while a high-brightness region with the same radius was centred 
to the south-east of \sgr\ (`east'), including the bright cloud G~0.11--0.11. At intermediate brightness 
values, a region of 102$''$ radius was selected that includes the Arches cloud and cluster (``Arches'').

For each time period and each region, the spectra of all observations and from all instruments have been fitted 
simultaneously with the model also used to estimate the continuum emission underlying the \six\ line in the imaging 
analysis (Sect.~\ref{section:analysis}), in order to have fully comparable results.  Considering that this model is 
not necessarily adapted to each  region, we find a good agreement between the two methods, with an average difference 
 of the imaging versus spectral fluxes of $3\%$, $-10\%$, $8\%$, and $-2\%$ for the north, south, east, and Arches 
 regions, respectively  (Fig.~\ref{Fig:ima_spe}). In particular, the most extreme outlier for the Arches region 
 corresponds to the 2006--2007 period  when the increased  \seven\ line due to a flare of the Arches cluster contaminates 
 the \six\ line in the spectral fitting \citep{capelli11}. 

 About 60\% of the tested regions have their \six\ fluxes consistent within 10\% with the spectral extraction and all but one are consistent
 within 15\%. Since the level of background emission is also varying with longitude in the CMZ, we also compared the fluxes
   in more distant regions. In the region G0.66-0.03 ($\ell \sim 0.7^\circ$, see Table~\ref{tab:regions}), the fluxes
   are consistent within 7\%, while in the weak and distant region Sgr D-Core ($\ell \sim 1.1^\circ$, see Table~\ref{tab:regions}),
   they are consistent within 16\%.
 We therefore conclude that our systematic uncertainty on the flux value resulting from imagery is at the level of 10--15\%. 

 We also evaluated the impact of the chosen plasma models to extract the 6.4 keV line flux. In particular,
   changing the abundance of the $kT=6.5$ keV apec to 2 as was found by, for example,  \citet{capelli12}, we found typical flux variations
   inferior to 3\% well below the level of systematics discussed above.


\begin{figure*}[!t]
\centering
\includegraphics[width=1.05\linewidth]{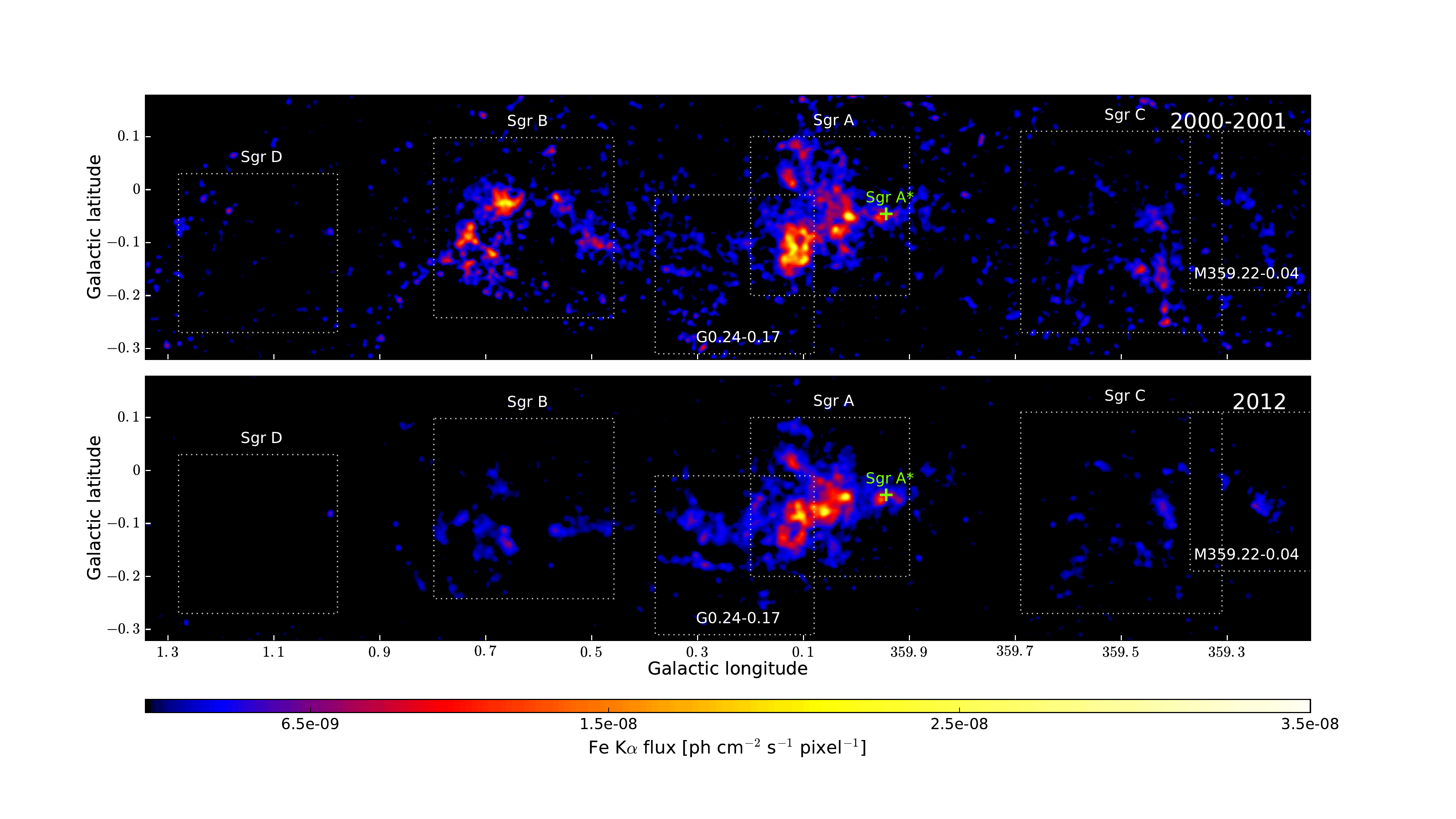}
   \caption{Background- and continuum-subtracted intensity maps of the inner GC region measured by \xmmN\ at 6.4~keV in 
        2000--2001 (top) and 2012 (bottom). The maps are in units of $\rm ph \, cm^{-2} \, s^{-1} \, pixel^{-1}$, 
        with 2.5$''$ pixel size, and smoothed using a Gaussian kernel of 5 pixels radius. The dotted square regions
        are discussed in more detail in Sect. \ref{section:variability_regions}.
        }
     \label{Fig:scans}
\end{figure*}

\section{Variability of the \fe\ line}\label{section:variability_general}

Thanks to the extensive \xmmN\ coverage of the CMZ we are for the first time able to follow the variations of the \fe\ 
line emission over 13 years simultaneously in several molecular complexes.  Figure~\ref{Fig:scans} presents the \fe\ intensity maps
of the 2000--2001 and 2012 \xmm\ scans (which cover similar sky areas) with 
a pixel size of 2.5$''\times 2.5''$.
The variability of the \six\ emission is clearly visible all over the central degree.
To quantitatively study variability on large scales and to produce light curves of specific
regions we used maps with a coarser binning of 1'$\times$1' pixels, which provide a larger number of counts per pixel. 

 We first compare the 2000-2001 and 2012 scans; then using all mosaics we determine regions of the CMZ which have \six\
 emission significantly varying over the 12 year period considered here.
 Finally in Sect. \ref{section:variability_regions}, we provide light curves of the \fe\ emission and discuss its evolution
 for most of the \six\ bright structures in the CMZ. 

\begin{figure}[!t]
\centering
\includegraphics[width=0.95\linewidth]{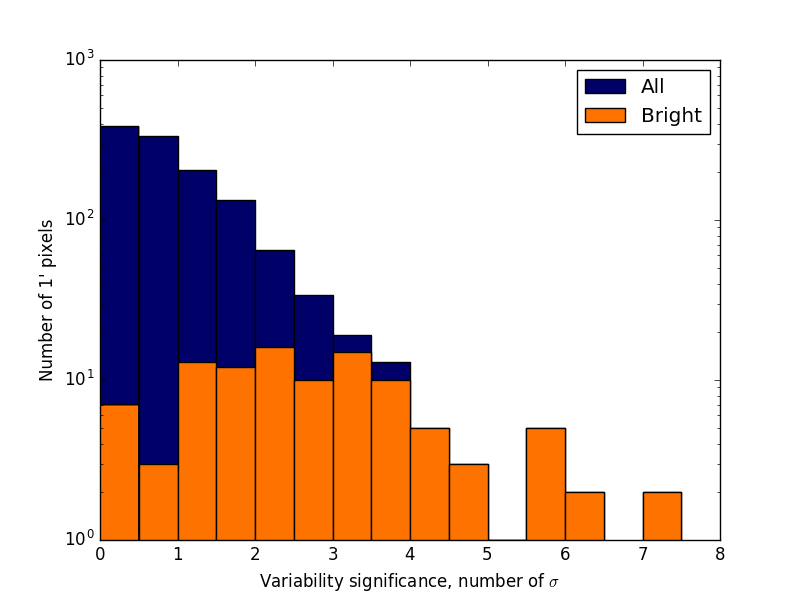}
\caption{Histogram of the significance of variations of \fe\ line emission in 1$'$ pixels between the 2000-2001 and 2012 surveys.
  Nearly all pixels with a brightness larger than $3 \times 10^{-6} \rm \, ph \, cm^{-2} \, s^{-1} \, arcmin^{-2}$ are found to vary.
}
     \label{Fig:histo}

\end{figure}

\subsection{Comparison of the 2000--2001 and 2012 scans of the CMZ}\label{section:var_scan}
When comparing the global \six\ maps obtained in 2000--2001 and 2012, it is already evident at  first glance that a global
decrease is observed between these two epochs \citep[see Fig.~\ref{Fig:scans} and][]{2014IAUS..303..333P}.
The total \six\ emission within a common rectangular
area of $19 \times 112 \rm \, arcmin^2$ size decreased from
$F_{\rm tot} (\text{2000--2001}) = (1.76 \pm 0.03) \times 10^{-3} \rm \, ph \, cm^{-2} \, s^{-1}$ 
to $F_{\rm tot} (\rm 2012) = (1.507 \pm 0.009) \times 10^{-3} \rm \, ph \, cm^{-2} \, s^{-1}$, corresponding 
to a total decrease of 14\%, which is marginally significant given the typical systematic error on the \fe\ line
flux obtained with the imaging approach. The average surface brightness over the same area decreased from
 $B_{\rm tot} (\text{2000--2001}) = (8.6 \pm 0.1) \times 10^{-7} \rm \, ph \, cm^{-2} \, s^{-1} \, arcmin^{-2}$ to 
$B_{\rm tot} (\rm 2012) = (7.36 \pm 0.04) \times 10^{-7} \rm \, ph \, cm^{-2} \, s^{-1} \, arcmin^{-2}$. 
These averaged values may be smoothing out strong variations occurring over a small fraction of the region.

In order to quantify the significance of the variations in the different regions, we applied to the 1~arcmin 
intensity maps the false discovery rate ({\it FDR}) technique. This method (whose application to astrophysical 
data is described by \citealt{miller01}) allows us to compare data against a model hypothesis while a priori controlling 
the number of false rejections, rather than setting a confidence limit as is done in standard procedures.
 The {\it FDR} has the advantage of providing a comparable rate of correct detections but fewer false detections 
 than other common statistical methods. 

Within the rectangular area common to the two scans, we retained only 
 pixels that were exposed during both scans  and where the \six\ emission was detected at 
$\ge 3\sigma$ in at least one of the two epochs.  As a result, 1092 pixels are selected in which we can test for 
variability. A false rejection rate of 10\% has been chosen in order to obtain a good 
balance between the number of false and true detections. We also verified with a rate of 5\% that the 
regions detected to vary are substantially the same as for 10\%. Out of the 1207 pixels tested, 60 show
significant variability, representing 5\% of the total. Of the variable  pixels, the large majority (80\%)
are found to decrease from 2000--2001 to 2012.

While they represent a small fraction of the total number of pixels, these variable pixels are in fact
all among the brightest ones. If we limit the sample to the pixels having a brightness larger than
$3 \times 10^{-6} \rm \, ph \, cm^{-2} \, s^{-1} \, arcmin^{-2}$ in at least one of the two scans, we find that
the majority (70 out of 105) of the bright pixels vary significantly (beyond 2$\sigma$, see Fig. \ref{Fig:histo}).
The significantly variable pixels are therefore the brightest. The numerous pixels which do not show
significant variation are, of course, not necessarily constant.
We lack the sensitivity to observe variations in fainter pixels mainly because the first survey is not deep enough.
The relevant point here is that most, if not all, bright pixels do vary on a ten year timescale.

When limiting the comparison to the bright \fe\ pixels, a larger decrease by more than 30\% is detected,
the average surface brightness of these pixels varying from 
$B_{\rm bright} (\text{2000--2001}) = (46.9 \pm 0.9) \times 10^{-7} \rm \, ph \, cm^{-2} \, s^{-1} \, arcmin^{-2}$ to 
$B_{\rm bright} (\rm 2012) = (31.7 \pm 0.3) \times 10^{-7} \rm \, ph \, cm^{-2} \, s^{-1} \, arcmin^{-2}$.

\subsection{Multi-year variability}\label{section:multi}

\begin{figure*}[!t]
\centering
\includegraphics[width=\linewidth]{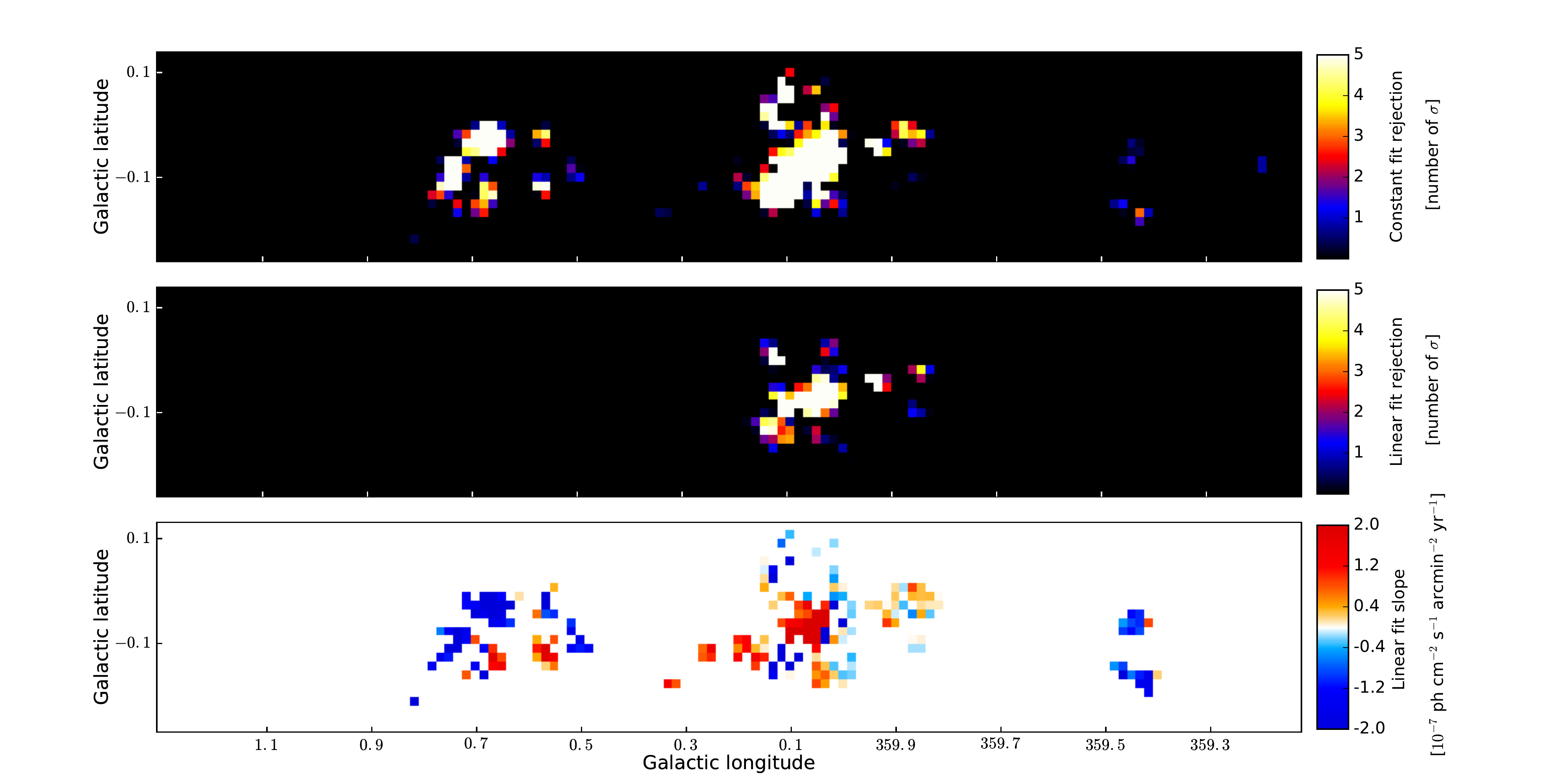}
   \caption{Maps resulting from the variability study of 7 periods from 2000 to 2012, using a 1$'$ pixel size.
            \textit{Top}: significance of rejection for a fit with a constant-intensity model applied to each bin 
            light curve, and after performing a $\chi^2$ smoothing (see text for details).
            \textit{Centre}: same as top panel but applying a linear fit.
            \textit{Bottom}: slope of the linear fit for the pixels for which variability can be detected and
             with non-zero significance shown in the top panel.
        }
     \label{Fig:var_map}

\end{figure*}

To characterise localised variations in the whole CMZ region over several years, we performed a systematic analysis
 following the method presented by \citet{clavel13}. We made use of the \fe\ mosaic images grouped by periods 
 of 2 years and extracted a light curve for each pixel of 1$'$ size (see Sect.~\ref{section:analysis}). 
 Each pixel's light curve was fitted with both constant and a linearly varying intensity model using  $\chi^2$  statistics
 to provide rejection probabilities for both models.  These resulting probabilities were then combined on scales of
 $2 \times 2 \rm \, arcmin^2$. The latter step has the advantage of emphasising the results on more
 extended regions showing similar trends, while on the other hand it makes it difficult to estimate a precise fraction 
of variable regions over the whole survey area, as a one-to-one correspondence between the initial maps and the 
significance map is lost.  To limit false variability detections, we corrected the probabilities for the number of trials,
which is given by  the total number of pixels in the map for which variability can be
detected\footnote{We considered that variability 
can be detected in pixels whose light curves have 1) at least 2 degrees of freedom for the constant fit and at least 3 for 
the linear fit; 2) at least one data point detected at more than $3\sigma$; 
3) a minimum detectable absolute slope for a linear fit 
$\le 0.7 \times 10^{-7} \rm \, ph \, cm^{-2} \, s^{-1} \, arcmin^{-2} \, yr^{-1}$, 
where the minimum slope is the smallest between that determined by the largest error bars of the data points in the light
 curve and that provided by the error bars of the two most distant points (in time) in the light curve.}. 
In Fig.~\ref{Fig:var_map} the results of the variability analysis are presented for this selection of pixels 
with sufficient statistics to be able to detect a minimum level of variability.

A significant deviation from a constant fit is observed in several large regions, which are associated
with the Sgr A, B, and C complexes. The structures in \sgrb, as well as most of the \sgra\ complex, present
the most significant variations ($\ge 4\sigma$ post-trial). The large majority of the variable pixels can
be clearly associated with either well-known bright molecular clouds, or to fainter molecular clumps. 
We note that the variable pixels just to the Galactic west of \sgr\ are most likely due to residual variability of the neutron 
star X-ray binary AX~J1745.6-2901, not completely removed during the exposure cut (see, e.g. \citealt{2015MNRAS.446.1536P}).

In order to have an indication of whether an increase or a decrease of emission is observed, the map giving the slope of 
the linear fit at each position is reported in Fig.~\ref{Fig:var_map} (bottom panel). Although the observed variations are, 
for the large majority, linear, non-linear variability has been significantly detected on arcminute scale in 
the \sgra\ complex (see centre panel of Fig.\ref{Fig:var_map}), in particular  in MC1 and MC2, in the Bridge,
around the Arches cluster, and in the south-east part of G0.11--0.11 (the regions are detailed in the Appendix
Table \ref{tab:regions} and in Sect. \ref{section:variability_regions:SgrA}).
This confirms the results obtained with \chandra\ observations from 1999 to 2011 on smaller scales for the Bridge as well as 
MC1 and MC2 clouds \citep{clavel13}.
Therefore, on the observed scales, there does not seem to be strong indication for non-linear variability except for
that already reported in the \sgra\ complex by \citet{clavel13}.
This is likely due to the more limited time sampling outside \sgra\ (see Fig. \ref{Fig:expo}).

In some regions, the variability is not significantly detected by this kind of analysis on arcminute scales;
for example, the northern part of the \sgrc\ complex and the G0.24--0.17 filamentary region (see later). This is due to the low surface brightness
of these regions and to the fact that propagation of the emission is observed here rather than an on-off behaviour.
 Indeed, when applying the same kind of analysis but combining the  $\chi^2$  probabilities on $3 \times 3 \rm \, arcmin^2$ 
 scales, the significance of rejection of a constant fit increases to $\sim 3\sigma$ for the northern region of \sgrc.
 However it still remains negligible for G0.24--0.17 because of its elongated structure.

\section{Variability in specific regions}\label{section:variability_regions}

We discuss here variability of specific regions. We consider most of the regions that have already been
the subject of a number of studies as well as a few others revealed in this analysis.
We refer to Table \ref{tab:regions} in the Appendix for details and relevant references. 
All errors shown on the light curves are 1 $\sigma$ errors (68\% confidence level).
\subsection{\sgra : two distinct timescales}\label{section:variability_regions:SgrA}

\begin{figure*}[!t]
\centering
\includegraphics[width=0.8\linewidth]{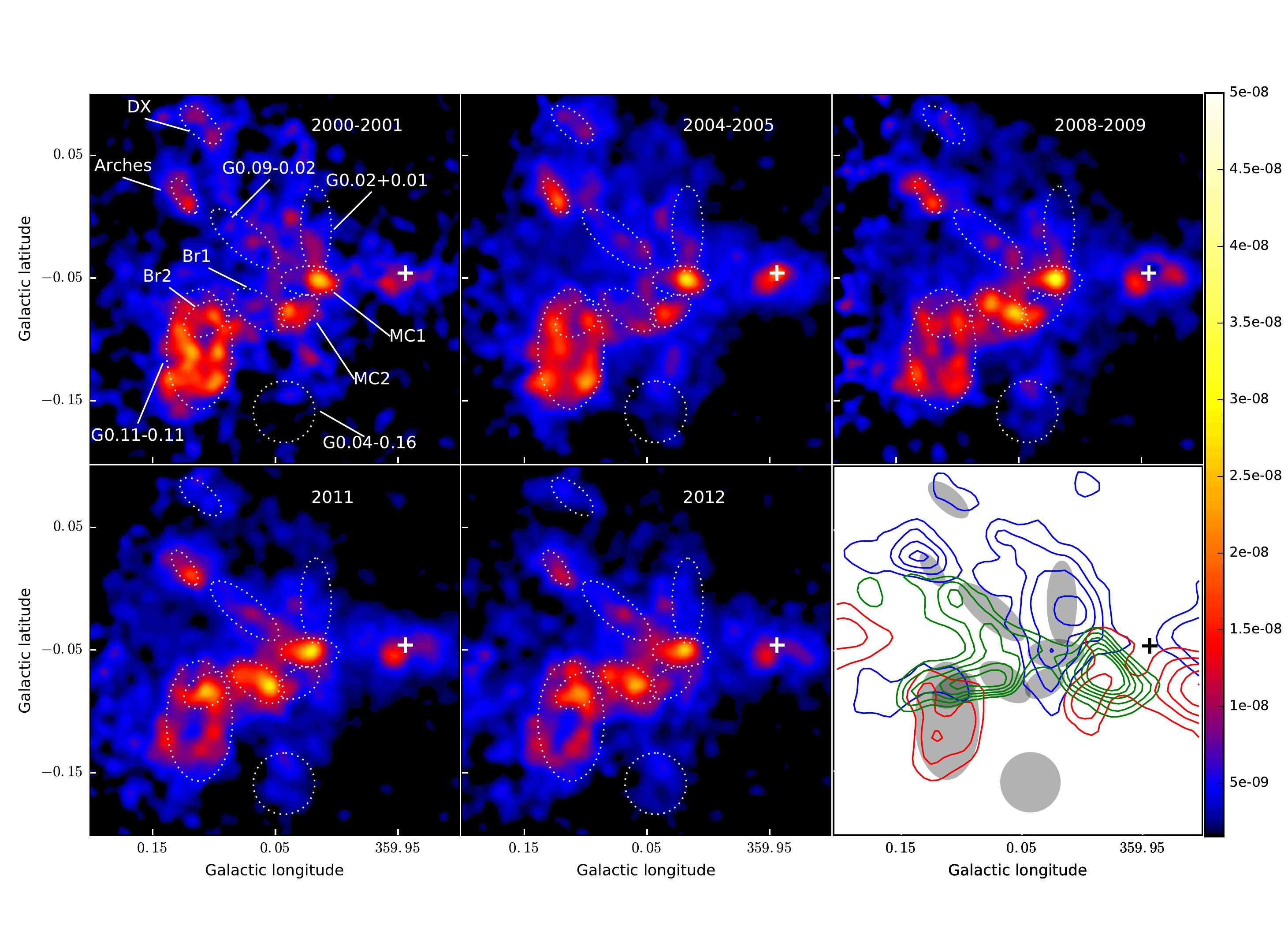}
   \caption{\fe\ flux images of the Sgr A complex in the period 2000-2012. The dotted lines mark the regions discussed in the
     text and whose light curves are extracted. The bottom right panel shows the same regions, as grey ellipses, compared to the
     various structures of the molecular complexes traced by CS shown as solid lines (blue -20 -- 10 km/s, red 10 -- 40 km/s
     and green 40 -- 70 km/s). Short timescale ($\lesssim$ 2 yr) variations are mostly visible in structures in the last
     velocity range.
                }
     \label{Fig:var_sgrA}
\end{figure*}

\begin{figure*}[!t]
\centering
\includegraphics[width=\linewidth]{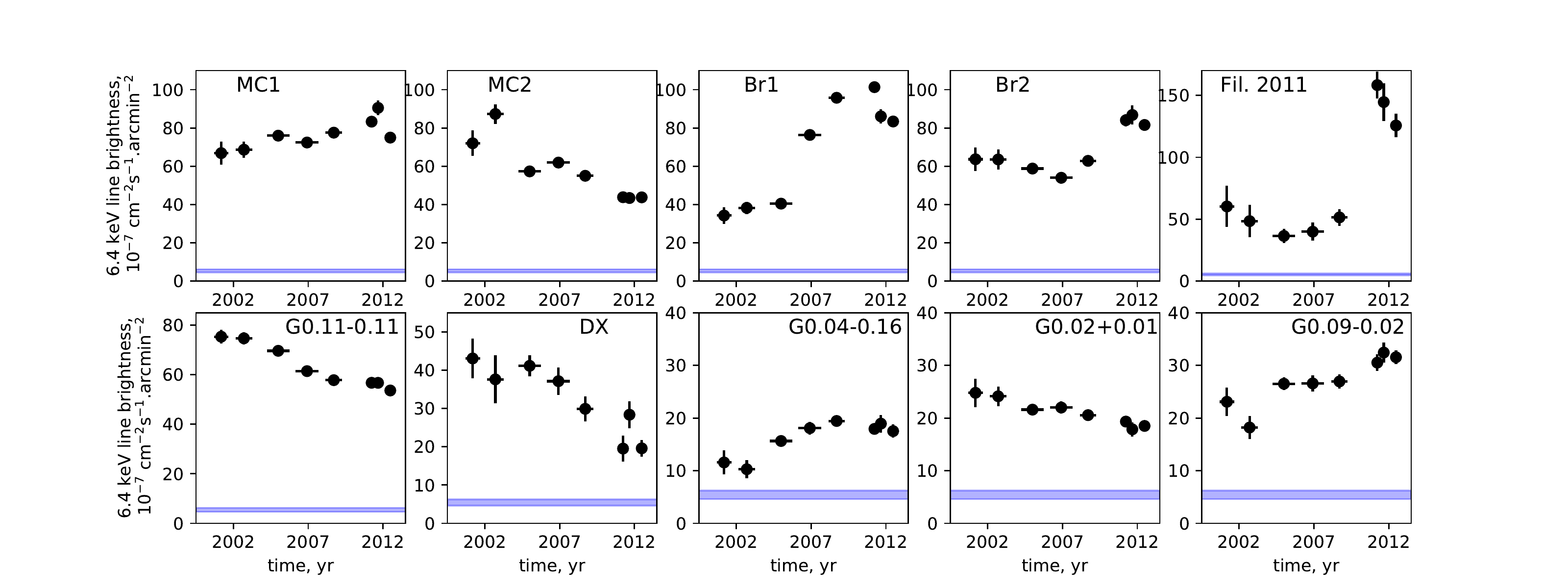}
   \caption{\fe\ light curve of the illuminated clouds in the Sgr A complex (regions indicated in Fig.~\ref{Fig:var_sgrA}). 
     The blue band shows the level of background line emission estimated in a large region below \sgr.
     All errors shown on the light curves are 1 $\sigma$ errors (68\% confidence level).}
     \label{Fig:lc_sgrA}
\end{figure*}


Figure \ref{Fig:var_sgrA} shows the evolution of the \fe\ emission in the whole \sgra\ complex. If the global
flux does not vary significantly, variability on small scales is very strong. Such variability 
has previously been detected in several regions with \xmm\ and \chandra\ \citep{muno07,ponti10,capelli11,clavel13, clavel14b}. Since the complete
\xmm\ data set covers a larger sky area and longer time period, we are able here to follow the evolution of known varying regions
and detect variability from additional molecular structures.

In particular, a complex variability behaviour is detected north of the MC1 and bridge clouds, in a region corresponding
to two molecular filaments that seem to connect these regions to the Arches and to the DX clouds.
These filamentary structures correspond to molecular matter in the -35 to 5 $\rm km \, s^{-1}$
and 35--65 $\rm km \, s^{-1}$ ranges, respectively \citep{tsuboi11}, and therefore similar velocity ranges
as the MC1 and Arches cloud for the former and the Bridge (Br1 and Br2 regions) for the latter (see Fig.~\ref{Fig:var_sgrA}).
The Arched filaments show globally increasing emission, while the \six\ emission of the western filament is decreasing
(see Fig.~\ref{Fig:var_map}). We extracted the light curves of these regions in two elliptical regions, named
G0.02+0.01 and G0.09-0.02 (see Table \ref{tab:regions}). The former shows a steady ~25\% linear flux decrease over the 12 years,
while the flux of the latter increases by more than 35\% in the same time period with a constant flux rejected at nearly 5 $\sigma$
(see last panels of Fig.~\ref{Fig:lc_sgrA}).

The Arches and DX clouds themselves are detected to have decreased during the 12 years separating the two surveys. 
A spectral analysis of the Arches region variability was performed and described in \citet{clavel14b}. Our new results 
are in full agreement and we refer to this paper for a light curve of the region. The DX light curve (Fig.~\ref{Fig:lc_sgrA}, last row second panel) shows a clear smooth decrease
(rejection of a constant fit at 6.4$\sigma$) and no indication for a peak in 2004 is found, contrary to what was reported
by \citet{capelli11}. To ensure that the discrepancy is not caused by the different time binning of 
the data, we extracted a light curve using only the observations and instrument exposures used by \citet{capelli11} and 
grouping them in the same way. We found flux points consistent with those obtained by \citet{capelli11} within the quoted 
uncertainties, but the overall light curve is smoother with no hint of a peak in August/September 2004, and thus not supporting the 
hypothesis of a strong flux increase in 2004.

On the other hand we can confirm that not only is the propagation of the signal, which has already been observed in 
several regions of \sgra, still on-going in the Br1, Br2, MC1 and MC2 clouds,  but that it is also observable in
other regions. In particular, the emission displacement is visible further away from \sgr\, , that is, within the 
G0.11--0.11 cloud,  where the two brightest regions of 2000--2001 appear to be shifted towards the Galactic east  and 
to have significantly dimmed in the 2012 map (Figs.~\ref{Fig:scans} and \ref{Fig:var_sgrA}). Moreover, we detect for the first time here 
the variability at \six\ of the region south of Br1/MC2, named G0.04--0.13 in \citet{clavel13}.
The light curve extracted from this region over the 12 years shows a smooth increase and a plateau (Fig. \ref{Fig:lc_sgrA},
last row middle panel). When looking on smaller scales (Fig.~\ref{Fig:var_sgrA}), one can observe
the propagation of the \six\ emission along a structure extending towards the south of the Bridge. The
CS emission of the molecular counterpart of this region is faint and is not visible in the last panel of Fig.~\ref{Fig:var_sgrA}.

As predicted by \citet{clavel13}, the large-scale emission of MC1 has started to decrease, following a peak in 2011. 
Within the elliptical region of Fig.~\ref{Fig:var_sgrA}, the flux in 2012 has decreased by about 15\% with respect 
to that in 2011 and the 12 year light curve now reveals significant variability in this cloud even 
on arcminute scales (Fig.~\ref{Fig:lc_sgrA}). A very thin filament ($0.2 \times 8 \rm \, pc^2$) was detected 
within Br2 in the 2011 \chandra\ observations during which it showed a sharp emission increase consistent with
a two-year variation \citep{clavel13}. It is also  detected by \xmm\ in 2011 and its flux is observed to be decreasing
in 2012 (by about 20\% compared to 2011),  supporting the hypothesis that it is reflecting a short flaring event (rejection
of a constant fit at 13$\sigma$;  Fig.~\ref{Fig:lc_sgrA}, first row, right panel). The corresponding region is not
shown in Fig.~\ref{Fig:var_sgrA} but is detailed in Table \ref{tab:regions}.

Molecular features highlighted by CS emission in the 40--70 km/s range all show strong variability with a typical timescale
of about 1--2 years (see the Br1, Br2 and 2011 filament light curves; the case of the faint region G0.09-0.02 is less clear).
As shown by \citet{clavel13}, they must have been illuminated by a short event with a maximum duration of $\sim$ 2 years.
All the other \six\ emitting regions display a smooth evolution over the 12 years covered and are likely due to
a longer-duration (typically 10 years) event.

\subsection{\sgrb: A strong flux decay and newly illuminated structures}

\begin{figure*}[!t]
\centering
\includegraphics[width=\linewidth]{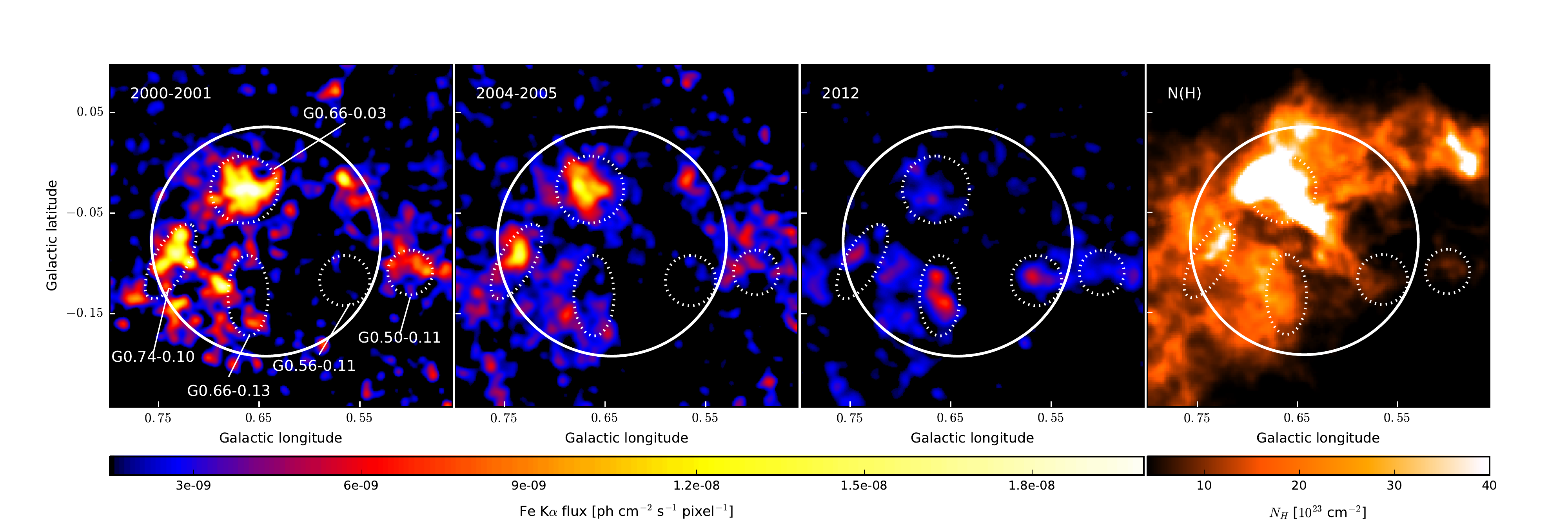}
\caption{\fe\ images of the \sgrb\ complex. The right panel shows the total gas column density measured by \textit{Herschel}
  \citep{2011ApJ...735L..33M}.} 
     \label{Fig:img_sgrb}
\end{figure*}

\begin{figure*}[!t]
\centering
\includegraphics[width=\linewidth]{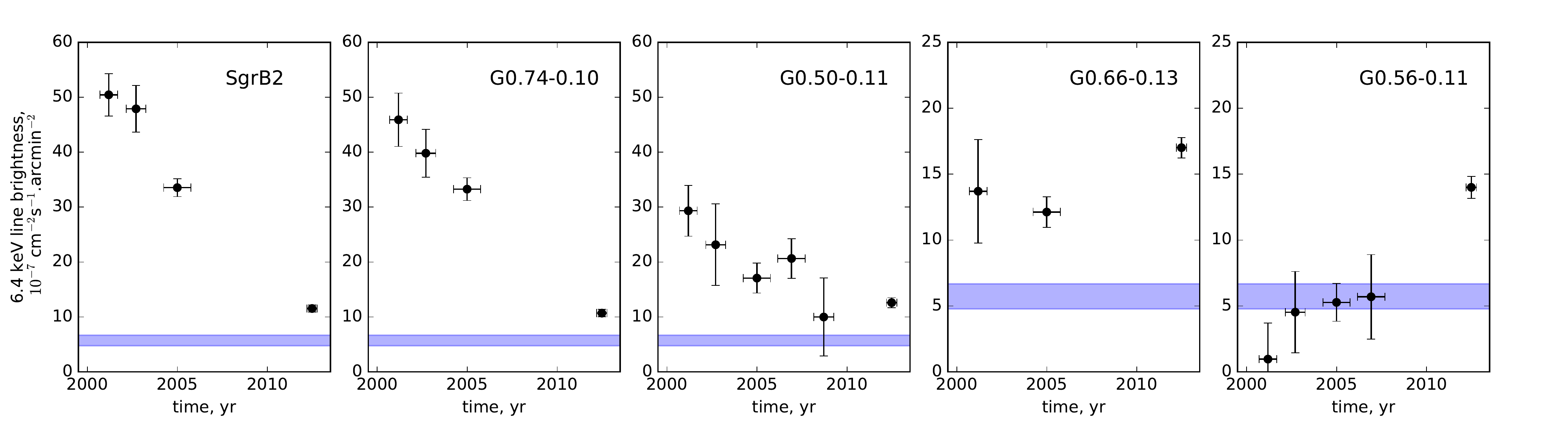}
\caption{\fe\ light curves of the \sgrb\ complex integrated over the regions indicated in Fig.~\ref{Fig:img_sgrb}.
  All errors shown on the light curves are 1 $\sigma$ errors (68\% confidence level).
        }
     \label{Fig:lc_sgrb}
\end{figure*}

The reflected emission in the \sgrb\ region is known to decay both in the \fe\ line and hard X-ray continuum 
\citep{inui09, terrier10, nobukawa11, Zhang15}; Figures \ref{Fig:img_sgrb} and \ref{Fig:lc_sgrb} clearly confirm this.

On large scales, the Sgr B circular region, which has a radius of $\sim$ 16 pc at the GC distance enclosing most of the molecular
complex, experienced a \six\ line flux variation of a factor of 2 over approximately 10 years. This variation is strongly statistically
significant with a constant level rejected at 18$\sigma$. The total flux peaked in 2000--2001 at 
$B_{\rm Sgr \, B}(2000-2001) = (17.7 \pm 0.7) \times 10^{-7} \rm \, ph \, cm^{-2} \, s^{-1} \, arcmin^{-2}$; it quickly dropped
to $B_{\rm Sgr \, B}(2004) = (13.6 \pm 0.3) \times 10^{-7} \rm \, ph \, cm^{-2} \, s^{-1} \, arcmin^{-2}$ and then decreased
more smoothly to reach $B_{\rm Sgr \, B}(2012) = (9.0 \pm 0.2) \times 10^{-7} \rm \, ph \, cm^{-2} \, s^{-1} \, arcmin^{-2}$.
This emission is still a factor of 2 larger than the more diffuse emission that pervades the inner degree and is therefore
clearly connected to the \sgrb\ molecular complex.

This behaviour at large scales masks more complex evolutions in smaller, localised regions. In the early measurements, most of the flux
is due to a few very bright regions. The flux of these compact and dense cores decayed much more rapidly than the ten-year timescale observed
in the larger region. If one excludes those cores, the large-scale brightness variation is less spectacular, as expected given the very large
light-crossing time of the whole Sgr B2 complex. The small and bright cores having short light-crossing time, have emission light curves
more closely reflecting the illumination light curve.

 The 2 arcmin around the core of the Sgr B2 cloud were very bright in 2000--2001, with a brightness of
$B_{\rm Sgr \, B2\ Core}(2000-2001) = (50.4 \pm 3.8) \times 10^{-7} \rm \, ph \, cm^{-2} \, s^{-1} \, arcmin^{-2}$, which fell by
 a factor of 4--5 over a duration of approximately 11 years to
 $B_{\rm Sgr \, B2\ Core}(2012) = (11.5 \pm 0.6) \times 10^{-7} \rm \, ph \, cm^{-2} \, s^{-1} \, arcmin^{-2}$
, which is only slightly larger than the average brightness in the whole \sgrb\ region in 2012.

The huge column density of the core  could also completely screen \six\ emission as the echo propagates deeper into the core and
produces a more rapid evolution than the actual illumination duration \citep[e.g.][]{Walls16}. In particular, we note that \textit{NuStar}
observed significant hard X-ray emission from the dense cores in 2013, most likely because of multiple scatterings in very high opacity regions \citep{Zhang15}.

A similar rapid decay is observed in the G0.74--0.09 region \citep{nobukawa11} where the flux is diminished by a factor of about 5 from   
$B_{\rm G0.74-0.09}(2000-2001) = (45.9 \pm 4.8) \times 10^{-7} \rm \, ph \, cm^{-2} \, s^{-1} \, arcmin^{-2}$ 
to $B_{\rm G0.74-0.09}(2012) = (10.7 \pm 0.7) \times 10^{-7} \rm \, ph \, cm^{-2} \, s^{-1} \, arcmin^{-2}$
, again only slightly larger than the average brightness in the whole \sgrb\ region.
The flux of region G0.50--0.11, discovered by \citet{2008PASJ...60S.191N}, is also declined by nearly a factor of 3 to reach a similar flux level.

Finally, the 2012 survey saw the appearance of two new illuminated regions. The first one, G0.6--0.13, was first reported by \citet{Zhang15}.
It is clearly visible on the maps (see Fig. \ref{Fig:scans} and Fig. \ref{Fig:img_sgrb}) and its light curve (fourth panel of Fig. \ref{Fig:lc_sgrb})
shows an increase with a 3.1 $\sigma$ statistical significance level. \textit{NuStar} observations suggest it was already dim in 2013 \citep{Zhang15},
suggesting that a short duration event might be propagating in the region.
The second region G0.56--0.11 has a flux level consistent with the background, except in 2012 where it reaches
$(14.0 \pm 0.8) \times 10^{-7} \rm \, ph \, cm^{-2} \, s^{-1} \, arcmin^{-2}$, rejecting a constant flux by 7 $\sigma$.

Further observations and studies of variations of these regions will be necessary to constrain the number and duration of the events
propagating in the Sgr B complex.

\subsection{\sgrc : Decrease and displacement}
\asca\ and more recently \suzaku\ have shown that several regions within the \sgrc\ molecular complex 
emit at \six\ \citep{murakami01,nakajima09,ryu13}. Only limited variability
had been reported by \citet{ryu13} for the X-ray clump named C1 
between two \suzaku\ observations (+8\% at 2.9$\sigma$ level), and a possible positional shift compared to past
\asca\ images was suggested \citep{nakajima09}.

\begin{figure*}[!t]
\centering
\includegraphics[width=\linewidth]{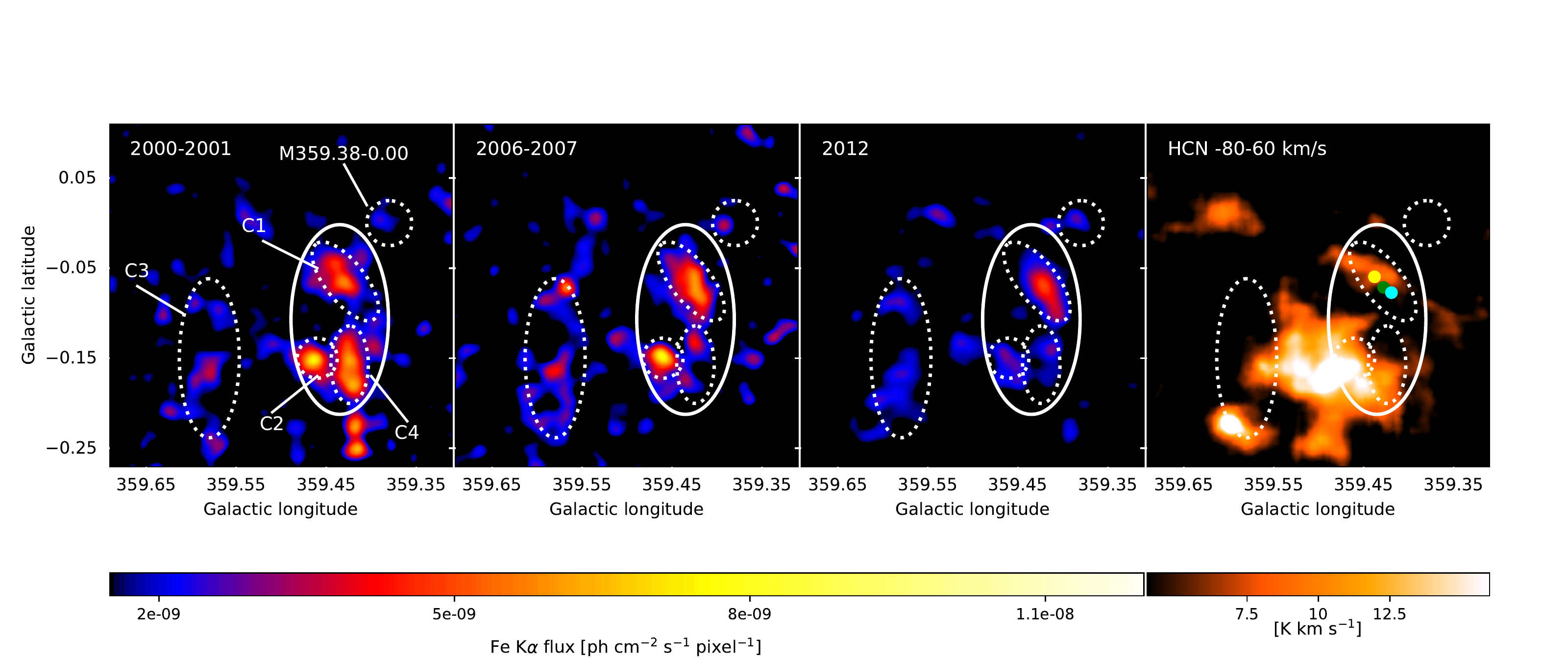}
   \caption{\fe\ maps of the \sgrc\ complex for three periods of \xmmN\ 
   observations (first three panels from the left) and Mopra HCN map integrated over the $-80$ to
   $-60 \, \rm km \, s^{-1}$ velocity band for the same region (right panel).
   The yellow, green, and cyan dots give the position of the centroid of the emission in the C1
     region  in 2000$-$2001, 2006$-$2007 and 2012, respectively. A displacement of 1.6' of the emission in
     a direction similar to that of the molecular feature is visible.
           The \fe\ maps are in units of $\rm ph \, cm^{-2} \, s^{-1} \, pixel^{-1}$, 
           with 2.5$''$ pixel size, and smoothed using a Gaussian kernel of 8 pixels radius.
           The solid line region labelled \sgrc\ includes the bulk of the \fe\ bright clumps, while 
           the smaller regions reported by other papers plus the C4 clump newly labelled here
           are shown in dotted lines. }
     \label{Fig:ima_sgrc}
\end{figure*}
\begin{figure*}[!t]
\centering
\includegraphics[width=\linewidth]{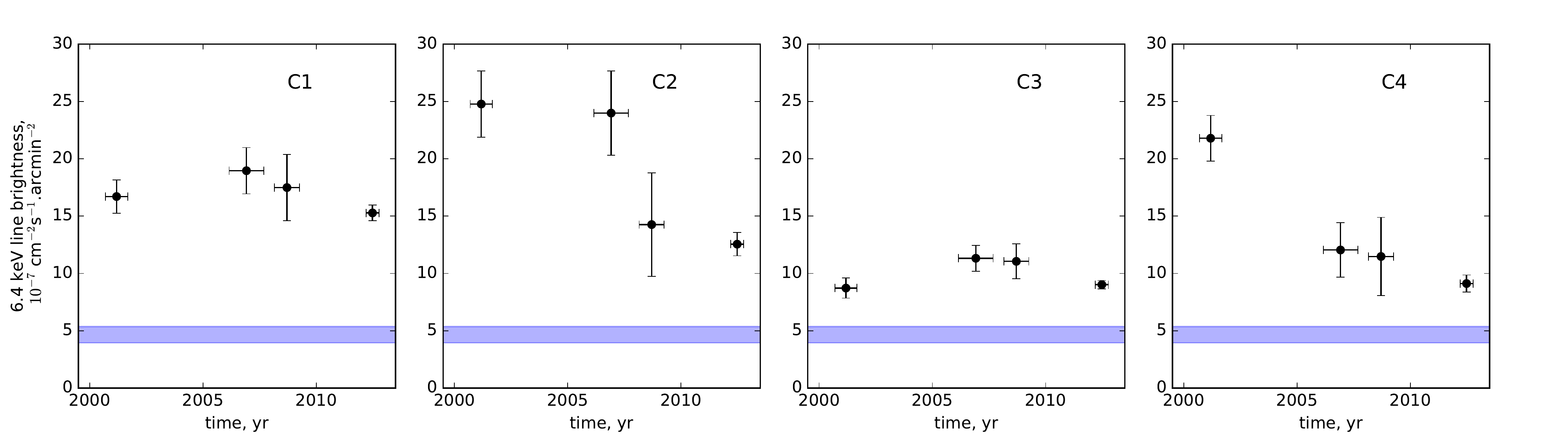}
   \caption{\fe\ light curves of the \sgrc\ complex (circles) integrated over the regions indicated 
   in Fig.~\ref{Fig:ima_sgrc}. The blue band indicates the level of background emission measured in a
   large region defined in Table \ref{tab:regions}.
   All errors shown on the light curves are 1 $\sigma$ errors (68\% confidence level).
        }
     \label{Fig:lc_sgrc}
\end{figure*}

The data presented here show for the first time a highly significant and large variation of the \fe\ flux 
in the \sgrc\ complex, detected with the same instruments and therefore free from cross-calibration issues.
When the bulk of the \six\ emission is integrated (elliptical region labelled \sgrc\ in Fig.~\ref{Fig:ima_sgrc}),
the total brightness is found to have decreased from 
$B_{\rm Sgr \, C}(2000-2001) = (12.7 \pm 0.6) \times 10^{-7} \rm \, ph \, cm^{-2} \, s^{-1} \, arcmin^{-2}$ to 
$B_{\rm Sgr \, C}(2012) = (9.1 \pm 0.3) \times 10^{-7} \rm \, ph \, cm^{-2} \, s^{-1} \, arcmin^{-2}$, 
that is, by 28\%. A constant flux is excluded at the 5.5 $\sigma$ significance level.
The amplitude of the variation is only a lower limit to the true amplitude, since the large-scale background 
level in the vicinity of \sgrc\ (and outside the bright clumps discussed here; see Fig.~\ref{Fig:ima_sgrc}) 
is on average about $B_{\rm bkg, \, Sgr \, C} \sim 6 \times 10^{-7} \rm \, ph \, cm^{-2} \, s^{-1} \, arcmin^{-2}$
(Fig.~\ref{Fig:lc_sgrc}).

The actual pattern of the variations is complex. Light curves of smaller-scale regions are presented in
Fig.~\ref{Fig:lc_sgrc}. These regions, described in Table \ref{tab:regions}, are taken from \citet{nakajima09} and \citet{ryu13}, and
we add a new one we call C4 following the nomenclature used in the latter paper.
We find no significant emission from the small clump M~359.38--0.00 
in excess of the background. The large region called C3 shows no significant variations.
Regions C2 and C4 both decrease significantly, with a constant flux rejected at 4.7 and 5.8 $\sigma$
, respectively. Their brightnesses have decreased by $\sim$ 50\% from 2000 to 2012.

The structure called C1 does not show a significant flux variation over the 12 years,
with a brightness measured at $(15.3 \pm 0.7) \times 10^{-7} \rm \, ph \, cm^{-2} \, s^{-1} \, arcmin^{-2}$ in 2012.
Yet, time-resolved images show a clear positional displacement of the \six\ centroid for the C1 clump
(Fig.~\ref{Fig:ima_sgrc}). To quantify the displacement, we determined the position of the centroid at each period
  by computing the barycenter of the emission in the C1 region from the flux maps shown in Fig.~\ref{Fig:ima_sgrc}.
In 2012, the centroid is displaced by about 1.6'  from the \sgr\ direction compared to its position in 2000--2001.
When comparing the centroid positions to the molecular matter distribution 
(rightmost panel of Fig.~\ref{Fig:ima_sgrc}), one can see the \six\ emission moving along a rather
filamentary structure, which is visible in the velocity range from --80 to --60 km/s in the HCN map.
Besides, while not perfectly aligned with the molecular matter, the \fe\ emission in C1 is elongated in
a similar direction. 
All this is suggests we are observing the propagation of the illuminated region along a dense matter
structure. This is the first time this phenomenon is observed at negative longitudes and it is consistent with
\sgr\ as the illuminating source. We can also use this propagation to constrain the duration of the illumination;
see Sect.~\ref{section:discussion}.


\subsection{Propagation in M359.23--0.04}

We also investigate the region M359.23--0.04 that was discovered by \citet{2010PASJ...62..971N}
during a \suzaku\ observation in 2008. It was found to be coincident with two molecular
features one appearing in the velocity range -140 to -120 km/s and a second in the range -20 to 0 km/s.
Because of its distance to \sgr\, \citet{2010PASJ...62..971N} argued that the origin of the illumination
could also be the nearby 1E 1740.7--2942 (6$'$ in projection).

\begin{figure*}[!t]
\centering
\includegraphics[width=0.75\linewidth]{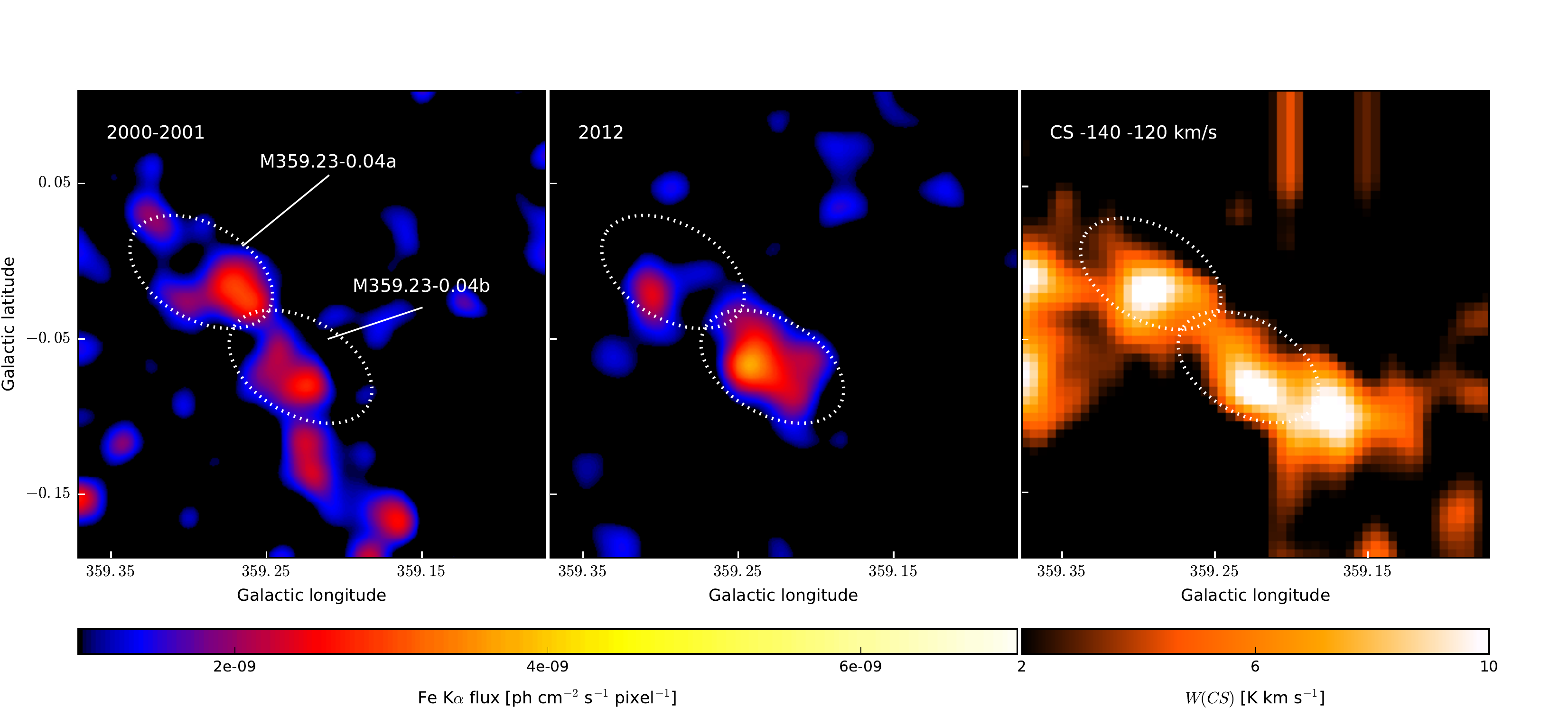}
\caption{\textit{Left images:} \fe\ maps centred at the position of the M359.23--0.04 structure during the 
   2000--2001 and 2012 \xmmN\ observations. The maps are in units of $\rm ph \, cm^{-2} \, s^{-1} \, pixel^{-1}$,
    with 2.5$''$ pixel size, and smoothed using a Gaussian kernel of 15 pixels in radius.
           \textit{Right:} CS(1--0) map integrated in the -140 to -120 $\rm km \, s^{-1}$ velocity band.
        }
     \label{Fig:lc_ima_G359}
\end{figure*}

M359.23--0.04 is visible only in our 2000-2001  and 2012 maps.
An elongated feature is visible and is found to be well aligned with the molecular matter traced by CS in the velo\-ci\-ty
band -140 to \mbox{-120 km/s} (see Fig. \ref{Fig:lc_ima_G359}). The latter is therefore the likely matter counterpart
of this reflection nebula. The nebula appears weakly variable in the systematic search presented in 
Sect. \ref{section:multi} but an apparent displacement of the emission towards the Galactic West is suggested by the  \fe\ maps (Fig. \ref{Fig:lc_ima_G359}).
This displacement follows the distribution of dense gas\footnote{We note that we have little exposure farther
  out and that we might miss emission from a fraction of the elongated matter structure. We note also that the
  morphology of the emission observed by \suzaku\ is very similar to that in our 2012 image.}. 

To quantify this, we define two regions encompassing the two blobs of the molecular structure which are well exposed with
  \xmm\ and we extract their \six\ flux (see Table \ref{tab:regions}).
The brightness of the first blob decreases from $(10.5 \pm 1.2) \times 10^{-7} \rm \, ph \, cm^{-2} \, s^{-1} \, arcmin^{-2}$
to $(6.6 \pm 0.6) \times 10^{-7} \rm \, ph \, cm^{-2} \, s^{-1} \, arcmin^{-2}$ (a 2.8 $\sigma$ variation),
while the brightness of the second, located farther away from \sgr, increases from  $(7.4 \pm 1.2) \times 10^{-7} \rm \, ph \, cm^{-2} \, s^{-1} \, arcmin^{-2}$
to $(12.1 \pm 0.9) \times 10^{-7} \rm \, ph \, cm^{-2} \, s^{-1} \, arcmin^{-2}$ in the same period (a 3.1 $\sigma$ variation).
 Although the statistical significance of the variations is not very strong, it is tempting to interpret this as the light echo
  propagating along a molecular structure, as in the case of the C1 feature discussed above. This positional shift towards the
  Galactic West would be consistent with an origin of the echo in the inner regions of the GC rather than in the nearby binary.
  In that case, as was shown by \citet{2010PASJ...62..971N}, a $\sim$ 10$^{39}$ erg s$^{-1}$ flare from \sgr\ could explain
  the observed luminosity. Further deep observations of the region are needed to convincingly validate this interpretation.

\subsection{\sgrd : Reflection in a distant cloud and weak evidence for variation}

Sgr D is located at the Galactic east of \sgr\ at a projected distance of about 180 pc and this more distant
molecular structure is also less massive than Sgr A, B or C. Therefore, this cloud is only intercepting a very
limited fraction of the signal emitted by the central source, making detection and variability studies of its
6.4 keV emission very challenging. Before our deep 2012 observation, Sgr D had been observed with \textit{BeppoSAX},
with \xmm\ (in 2000 and 2005) and with Suzaku (in 2007). These earlier observations revealed multiple sources of X-ray
emission in the region \citep{2001A&A...372..651S,2006A&A...456..287S,2009PASJ...61S.209S, Ponti15a}, including an extended 6.4 keV
clump \citep{ryu2013PHD}. In 2007, this extended feature had a surface brightness of
$I_{6.4keV} = (2.7 \pm 0.3) \times 10^{-7}$ ph cm$^{-2}$ s$^{-1}$ arcmin$^{-2}$ (unabsorbed, corresponding to a measured
brightness of $ \sim 2.2 \times 10^{-7}$ ph cm$^{-2}$ s$^{-1}$ arcmin$^{-2}$ and a possible X-ray nebula
origin was then mentioned \citep{ryu2013PHD}.

Making use of the 2012 observations covering Sgr D, we confirm this result with a clear detection of its 6.4 keV
emission with \xmm. The bulk of the \fe\ emission is contained in the elliptical region
we call ‘Sgr D core’ (see Fig. \ref{Fig:sgrd} and Table \ref{tab:regions} of the Appendix). This emission is found to be
coincident with a molecular feature in the CS data in the velocity range 50 to 70 km/s.
In 2012, it has a surface brightness of $3.8 \pm 0.3 \times 10^{-7}$ ph cm$^{-2}$ s$^{-1}$ arcmin$^{-2}$. 
In 2000 and 2005, the 6.4 keV line emission of this  region was compatible with the background level.
However, the lower statistics of these earlier observations only provide weak evidence for the variability of Sgr D:
the constant fit of the \xmm\ light curve is rejected at a 3.3 $\sigma$ level and a linear fit is preferred over
the constant one at a 3.7 $\sigma$ level\footnote{The ‘Sgr D core’ region we consider here is only partly included
in the larger region chosen for the Suzaku analysis \citep{ryu2013PHD}, so this additional 6.4 keV measurement in 2007 could
not be used in our variability analysis.}.

If this increasing trend is confirmed, Sgr D would be the furthest dense cloud from \sgr\ (in projection) for which the \fe\ emission and
variability are observed, providing a constraint on the past event from \sgr\ causing this echo to have occurred between 300 and 1100 years
ago (assuming that the line of sight distance of this cloud to the sky plane of \sgr\ is less than 130~pc).

\begin{figure*}[ht]
\centering
\includegraphics[width=0.6\linewidth]{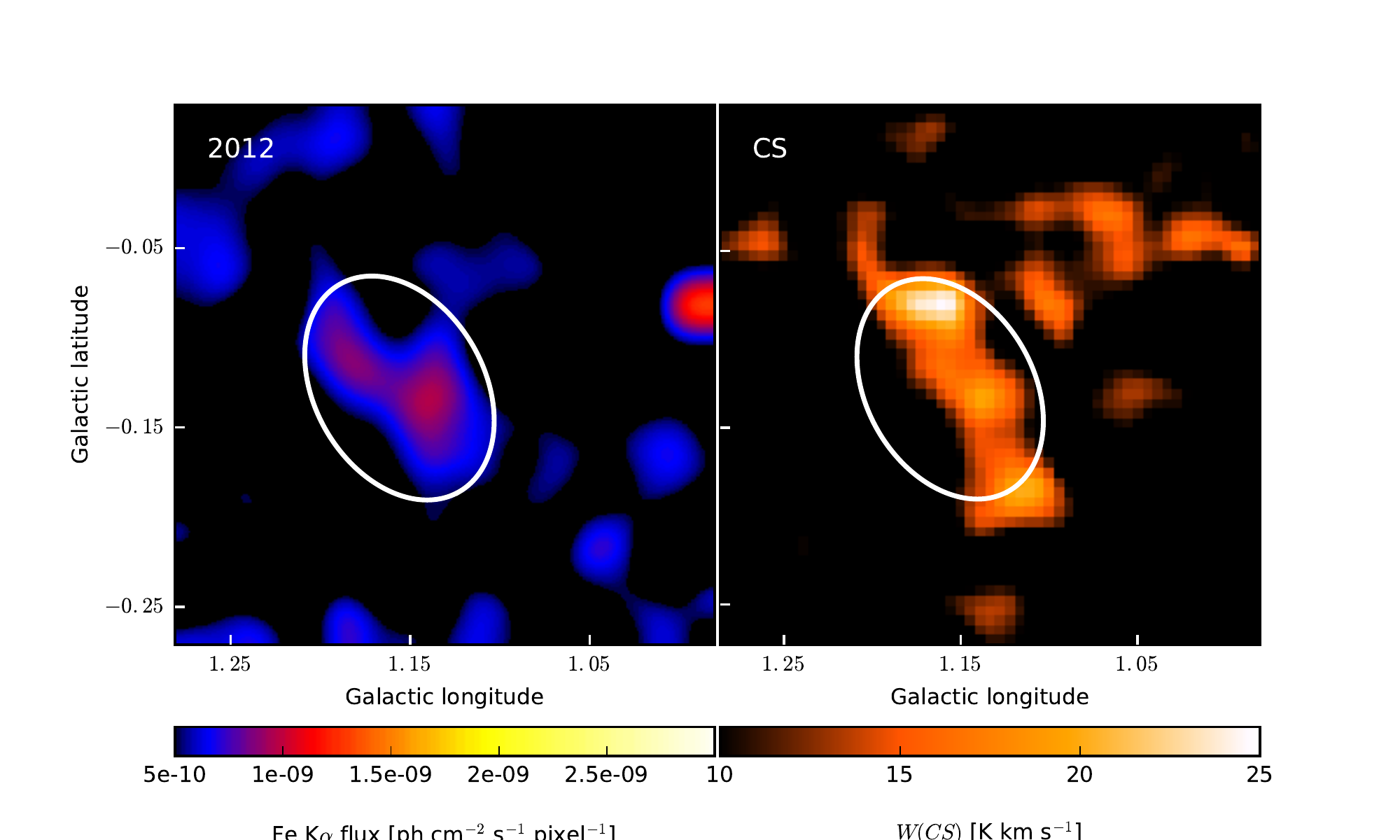}
\includegraphics[width=0.3\linewidth]{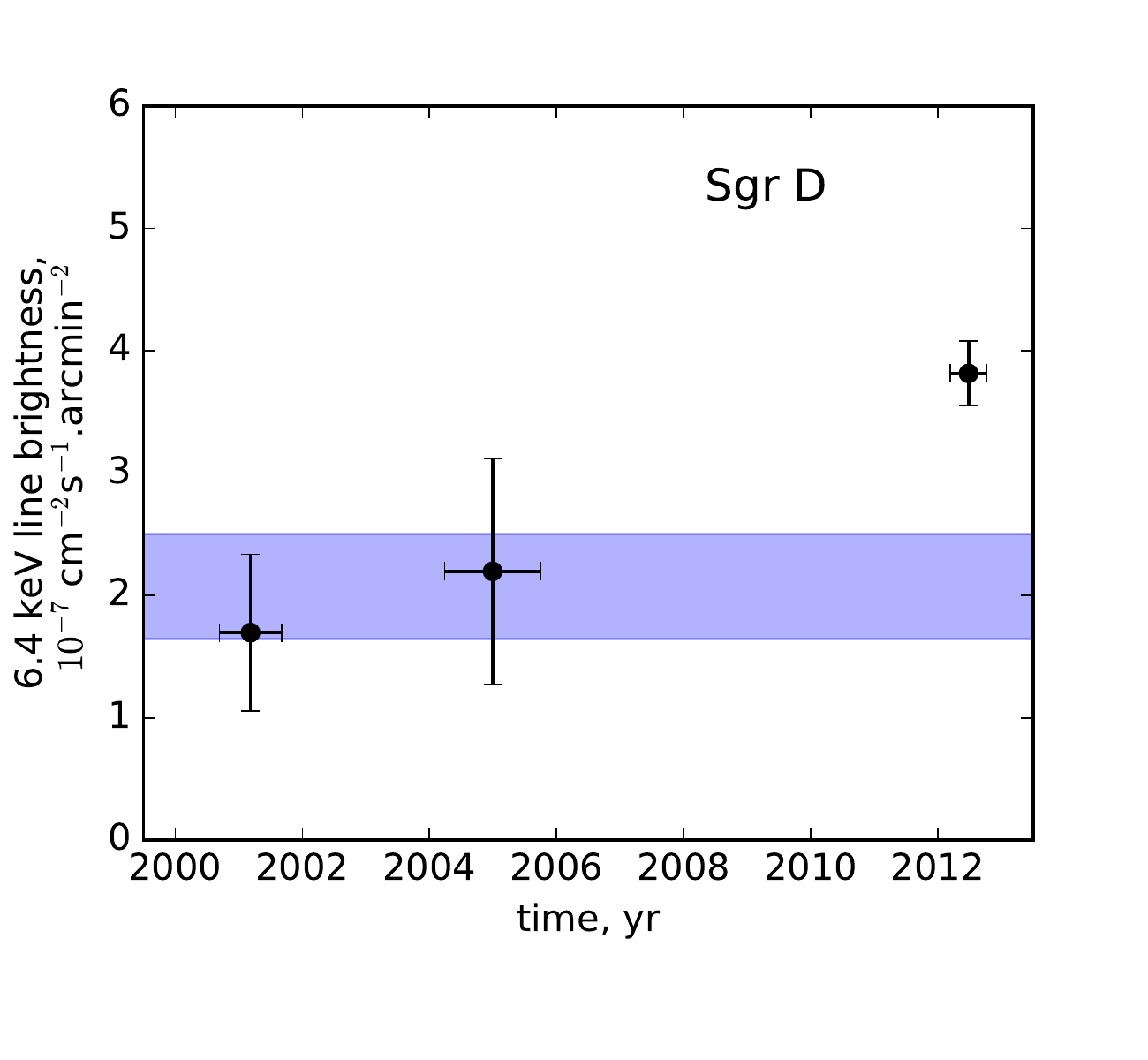}

\caption{(left) \fe\ map centred at the position of \sgrd\ during the 2012 \xmmN\ observations. The map is in units of
  $\rm ph \, cm^{-2} \, s^{-1} \, pixel^{-1}$, with 2.5$''$ pixel size, and smoothed using a Gaussian kernel of 15 pixels
  radius. The region labelled \sgrd\ includes the bulk of the \fe\ emission in the complex. (middle) CS map integrated in
  the velocity range 50 to 70 km/s showing the molecular counterpart of the \six\ emission. (Right) \fe\ light curve of the \sgrd\
  complex integrating over the region indicated on the maps. The blue band indicates the level of background emission.
  All errors shown on the light curves are 1 $\sigma$ errors (68\% confidence level).
}
\label{Fig:sgrd}
\end{figure*}


\subsection{G0.24--0.17: Rapid illumination of a 25 pc-long structure}

\begin{figure*}[!t]
\centering
\includegraphics[width=0.67\linewidth]{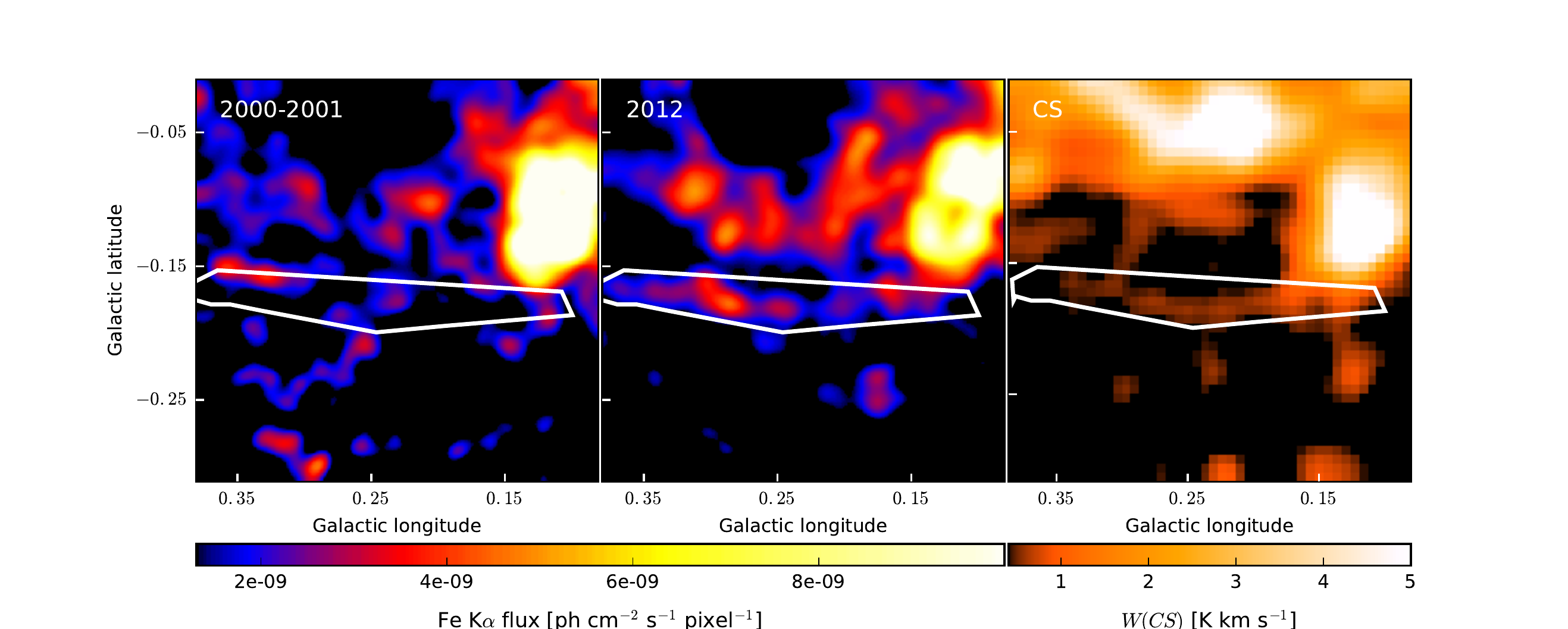}
\includegraphics[width=0.3\linewidth]{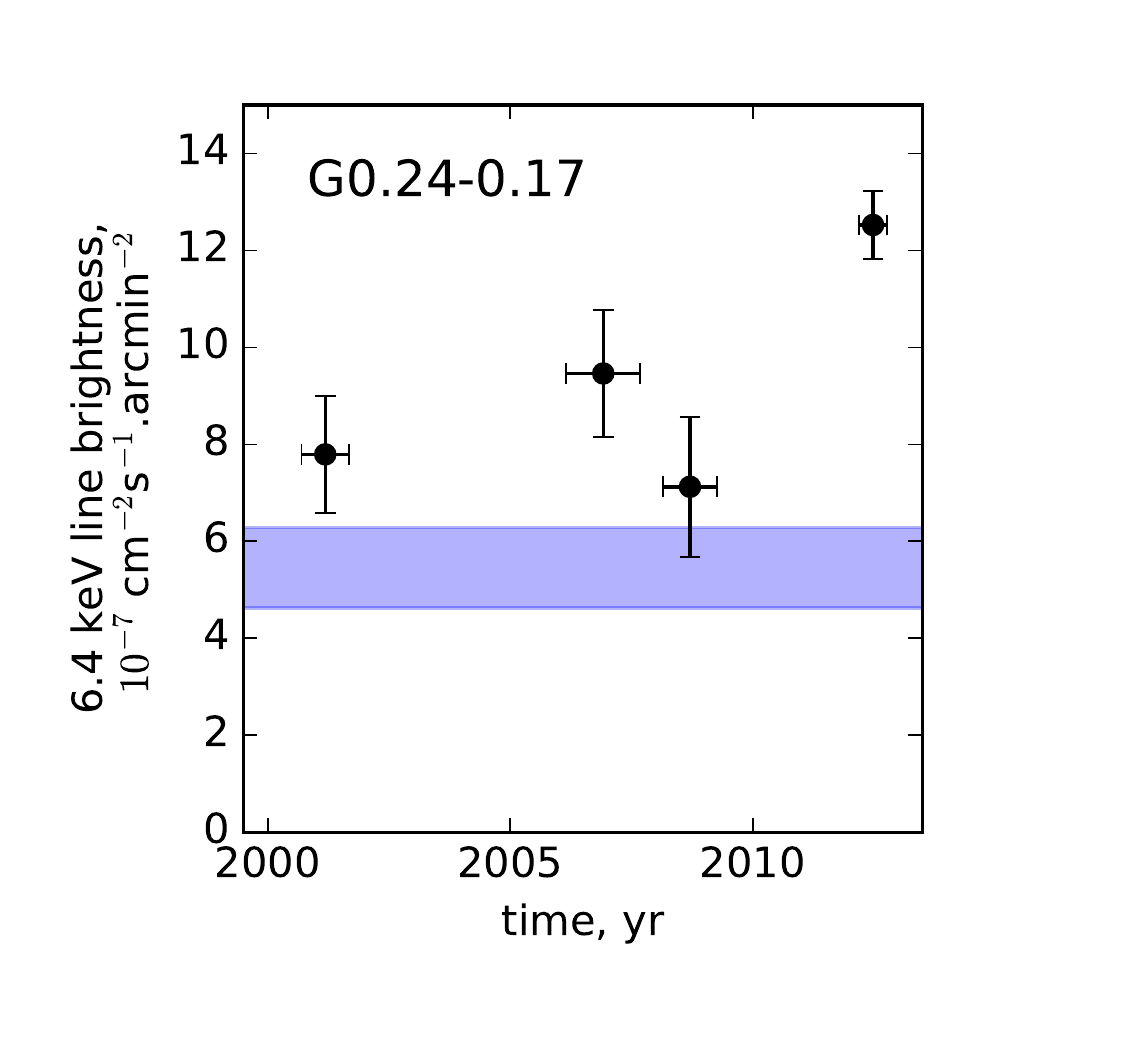}
\caption{\textit{Left images:} \fe\ maps centred at the position of the G0.24--0.17 filament during the 
   2000--2001 and 2012 \xmmN\ observations. The maps are in units of $\rm ph \, cm^{-2} \, s^{-1} \, pixel^{-1}$,
    with 2.5$''$ pixel size, and smoothed using a Gaussian kernel of 8 pixels in radius.
           \textit{Centre:} Mopra CS(1--0) map integrated over the 27--35 $\rm km \, s^{-1}$ velocity band.
   \textit{Right:} \fe\ light curve of the G0.24--0.17 filament.
   The blue band indicates the level of background emission measured in a circular region south of \sgr.
   All errors shown on the light curves are 1 $\sigma$ errors (68\% confidence level).
        }
     \label{Fig:lc_ima_filament}
\end{figure*}

An interesting filamentary structure that appeared to be newly illuminated in 2012 (hereafter G0.24--0.17) 
is observed south-east of the \sgra\ complex. This feature has an elongated shape, roughly parallel 
to the Galactic plane (Fig.~\ref{Fig:lc_ima_filament}) which connects to the G0.11--0.11 cloud on its
western edge.
Looking into molecular tracer maps to search for a corresponding feature, we found a relatively faint filamentary
structure of about $25 \times 2 \rm \, pc^2$ in projected size in the  CS(1--0) maps obtained from the 7~mm Mopra
survey of the CMZ \citep{jones13}, in the velocity range between 27 and 35 $\rm km \, s^{-1}$ (see Fig.~\ref{Fig:lc_ima_filament}).
While not significant in itself, the very clear morphological match between the \six\ and CS features strongly
suggests the latter is real and is the molecular counterpart of the \fe\ line emission observed by \xmmN\ in 2012.\\

We selected a polygonal region containing most of the CS line  and the 2012 \fe\ line emissions  of the filament
and computed the total flux at
different epochs (Fig.~\ref{Fig:lc_ima_filament}). A constant fit of this light curve is rejected at 3.9 $\sigma$.
For each epoch we also compared the surface brightness of the filament with that of the background within the same mosaic image,
estimated over a circular region south of \sgr, free of bright \six\ emission. In the 2012 mosaic, a total brightness of 
$(12.5 \pm 0.5) \times 10^{-7} \rm \, ph \, cm^{-2} \, s^{-1}\, arcmin^{-2}$ is measured 
compared to a 2000--2001 brightness of $(7.8 \pm 1.2) \times 10^{-7} \rm \, ph \, cm^{-2} \, s^{-1}\, arcmin^{-2}$, 
which is compatible with the average background  within 2$\sigma$ (Fig.~\ref{Fig:lc_ima_filament}).
In addition, the brightest emission in 2000--2001 seems to rather come from across the north-east border of the polygonal region.

This suggests that the illumination propagated through the 25 pc-long filament within less than the 12 years
separating the two surveys; another case of superluminal echo. This places strong constraints on the filament location and orientation with respect to
the illuminating source. Furthermore, the faintness of the CS counterpart indicates that the column density of this
molecular structure is small and hence that the luminosity of the source must be large. We discuss these constraints
further in the following Section.

Finally, we note that a complex region of \six\ emission is also visible in 2012 around $-0.10^\circ < b < -0.15^\circ$,
  and $0.15^\circ < \ell < 0.35^\circ$; see Fig.~\ref{Fig:lc_ima_filament}. The flux variation of this feature has positive slope
  in Fig.~\ref{Fig:var_map}. This suggests that the illumination is propagating farther than the Sgr A molecular complex.

\section{Discussion}\label{section:discussion}

\subsection{Timescales of variations}

We have shown, in Sect. \ref{section:var_scan}, that most bright regions have significantly varied over the 12
years separating the two \xmm\ surveys. In particular, very few regions that were bright during the 2000--2001 scan have a compatible flux
in 2012. We can therefore conclude that typical timescales in these regions are of the order of or shorter than 10 years.
Those regions are not illuminated by a century-long steady event; if they were we would have observed more stable flux levels.

We also confirm the existence of two distinct timescales in the variations observed in the Sgr A complex as was first noted
by \citet{clavel13}. These patterns are observed in distinct molecular structures
which are located in different velocity ranges. The rapid variations are limited to structures in the 40--70 km/s velocity range
(the so-called Bridge), while the longer variations (10 year timescale linear variations) are visible mostly in molecular
features visible in the -20--10 km/s and 10--40 km/s ranges (see Fig. \ref{Fig:var_sgrA}). If, as suggested by \citet{Churazov16},
every \fe\ bright cloud is illuminated by only one short event, they all should lie in a very limited range of positions
along the line of sight. This seems unlikely. Indeed, while one can assume that some of the clouds are interacting, we note that the molecular
structures, which represent most of the mass in this direction, are distributed over a broad range of velocities and cover more than 20 pc
in projected distance. Therefore we consider that several events are required to explain the observations.

The variations observed in the Sgr B complex are rather rapid. The bright regions, Sgr B2 core and G0.74--0.10,  have decayed
by a factor of 4--5 in 12 years and are now at a level similar to the average brightness level of the Sgr B complex.
This decrease is much more rapid than what is observed in most regions of the \sgra\ complex, such as MC1 and MC2, which have
similar or smaller physical sizes. 
This might suggest, at first, that these clouds are illuminated by a short flare of a few years in duration,
but absorption and propagation effects can also strongly modify the light curve. Sgr B2 is the most massive cloud in the Galaxy,
with a peak column density beyond $10^{24}$ cm$^{-2}$ \citep{2008MNRAS.390..683P} and is therefore opaque at \six.
In this energy range, we can only see reflection from the skin of the densest regions, whereas hard X-rays can still escape
from the cores, as seen by \textit{NuSTAR} \citep{Zhang15}. The \six\ light curve of less dense clouds in the region should follow more closely
the illumination. In that respect, we note that the future evolution of the newly illuminated features, G0.66--0.13 and
G0.56--0.11, will be very beneficial for constraining the illuminating event light curve. \textit{NuSTAR} observations carried out in 2013
suggest the flux of the former has already decreased \citep{Zhang15}.

\subsection{Duration of the illumination in Sgr C}

The overall emission in the Sgr C complex is decreasing. In particular, regions C2 and C4 show a significant
and rapid decrease. The case of region C1 is more subtle: if the total flux does not show any significant variation,
the centroid of the emission is moving in a direction following the elongated molecular structure shown in Fig. \ref{Fig:ima_sgrc}.
  We argued that this is consistent with the displacement of the illuminated region of the molecular structure because of the
  lightfront propagation. If this is correct, we can constrain the duration of the event reflected in the C1 region.

In 2000, the illuminated length was $\Delta l = 2.9'$ or about 6.5~pc  at the GC distance. In 2012, this illuminated length 
was found unchanged but the centroid of the emission was offset by $\delta x = 1.6'$.
From these measurements, we can extract an estimate of the duration, $\Delta T$, of the event illuminating this region.
A priori, the unknown position and inclination of the structure with respect to the line-of-sight hampers a direct estimation
of $\Delta T$. Yet, if we assume that the molecular structure 
is rectilinear,  we have $\Delta l = v_{app} \Delta T$, where $v_{app}$ is the apparent velocity of the light front 
along the linear feature. 

Similarly, if the time between two measurements, $\delta t,$ is short compared to the age of the event, it is proportional to 
the displacement along the structure, $\delta x = v_{app} \delta t$. We have then:

\begin{align}\label{eq1}
  \frac{\delta t}{\Delta T} = \frac{\delta x}{\Delta l} ,
\end{align}

where $\delta x = 1.6'$ and $\delta t = 12$ yrs. We deduce that $\Delta T \sim 22$~years.

The duration of the event illuminating the C1 region of the \sgrc\ complex is therefore comparable to the one observed
in the G0.11--0.11 or  MC1 clouds of the \sgra\ complex by \citet{clavel13}, which has a typical rise-time of 10 years.
It is therefore plausible that the event illuminating this part of \sgrc\ is also the one causing variations in \sgra.     
Constraints on the geometry will be necessary to test this hypothesis.

\subsection{Illumination of a 25 pc long feature in 10 years}

The newly illuminated filamentary region G0.24--0.17 is approximately 25 pc $\times$ 2 pc in projection. It is apparently fully
illuminated in 12 years or less. It must therefore have a very specific location and angle with respect to the line of sight. At any point,
the surface of equal delays is a paraboloid centred on \sgr, so the filament has to be nearly tangent to one such paraboloid.

Except for a 3.5$'$ long enhancement in the region of l=0.285$^{\circ}$, the 6.4 keV intensity is relatively flat along the filament.
Since the light curve of the flare should be partly reflected along the structure, this would indicate that there are limited
flux variations in the illuminating flux.

Given the very low signal in the CS data, the molecular filament must be thin. The integrated CS flux is more than an order of magnitude
smaller in the filament than that in the G0.11--0.11 cloud. The $\rm N_H$ estimated for the latter range from
$2 \times 10^{22}$ cm$^{-2}$ \citep{AmoBaladron09} to $6 \times 10^{23}$ cm$^{-2}$ \citep{Handa06}.  The $\rm N_H$ of the G0.24--0.17
filament can therefore not be larger than a few $10^{22}$ cm$^{-2}$. Here, we assume a density of about 1000 cm$^{-3}$ for the
structure; the total $\rm N_H$ integrated over its 3 pc thickness is then $\sim 10^{22}$ cm$^{-2}$. The cloud is therefore optically thin.

If we neglect flux dilution along the filament, we can apply the simple derivation given by \citet{Sunyaev98} to infer the typical
minimal brightness of the illuminating source. Because the cloud is not spherical here, we have to correct the formula for the apparent
size of the filament seen from the illuminating source. Remembering that the cloud has to be tangent to the paraboloid of equal delays
to be quickly illuminated, we can conclude that the apparent size of the filament seen from \sgr\ is the same as its apparent size for
the observer at infinity, because of the geometrical properties of the parabola. We therefore get the luminosity of the illuminating event: 
\begin{align}\label{eq2}
  L = 1.9 \times 10^{39} \text{erg s}^{-1} \left(\frac{N_H}{10^{22} \text{cm}^{-2}} \right)^{-1} \left(\frac{d}{46 \text{pc}} \right)^2 ,
\end{align}
where $d$ is the physical distance from \sgr\ to the centre of the filament. We assumed an $E^{-2}$ power-law spectrum to estimate
the luminosity over the 2 to 10 keV range. This value is consistent with values obtained for other clouds in the CMZ.

It is likely that the illuminating event is one of those visible in the \sgra\ complex and discussed in Sect.
\ref{section:variability_regions:SgrA} and in \citet{clavel13}. The apparent connection
with the G0.11--0.11 cloud, visible both at \six\ and in CS molecule line maps, suggests that the signal in the latter cloud
is now propagating in G0.24--0.17. Further monitoring of the filament will be necessary to see whether the time behaviour is
similar. A rapid decay of the flux would suggest a shorter event than the one observed in G0.11--0.11.


\section{Conclusions}\label{section:conclusions}

We have performed a systematic analysis of \fe\ line variations in the central 300 pc with \xmmN\ over a period spanning 12 years
from 2000 to 2012. In particular, two complete surveys of the region were performed in 2000-2001 and 2012 and allow a
comparison of variations of the line flux. We find that, in the limit of the surveys' sensitivity, most of the
\Fe\ bright regions show significant variations between the two dates. This excludes sub-relativistic ion bombardment as
the origin of the bright regions since their cooling time is significantly longer. We cannot exclude that cosmic rays
play a significant role in some fainter, constant, diffuse \six\ emission regions \citep[as discussed e.g. in][]{capelli12}.
Yet, we note that \citet{2013ApJ...771L..43D} have shown
that such a scenario applied to the overall diffuse \six\ emission in the CMZ would lead to ionization levels, as inferred
from H$_3^+$ molecule emission, much larger than observed.

The rapid variations of the \six\ bright regions are also evidence that the reflection consists of relatively brief
illumination episodes, lasting typically ten years or less. If a long duration steady illumination was significantly
contributing to the illumination of the inner 300 pc, we should have observed regions of bright and stable \fe\ emission. 
The conclusion of \citet{clavel13} that at least two distinct events, with typical durations of 2 and 10 years, are
required to explain the variations is confirmed. This is supported by the presence of variations on two distinct time scales
observed in the Sgr A complex, by the rapid variations observed in the Sgr B complex and by the measurement of
a displacement of the illumination in the Sgr C region. Whether more events are required is still an open question.
In particular, we note that the emission from the more distant Sgr D complex could have a significantly larger delay
and could be caused by a more ancient event.

Calculation of the exact date of the outbursts will require detailed spectral modelling of molecular clouds \citep{Walls16,Chuard17}.
Other approaches based on the apparent velocity
of reflected features would also allow one to date the illumination \citep{Churazov16} assuming it is caused by a single flash.  
Further monitoring of the illuminated regions will be crucial to disentangle illumination and matter distribution effects.
Long-term evolution of the reflection as well as X-ray polarimetric observations will also help constrain the location of
the illuminating source \citep{2016arXiv161200180C}.


\begin{acknowledgements}
      The authors acknowledge the Centre National d'Etudes Spatiales (CNES) for financial support.
      GP acknowledges support via an EU Marie Curie Intra-European fellowship under contract no. FP-PEOPLE-2012-IEF-331095.
      This work has been partly supported by the LabEx UnivEarthS\footnote{\tt http://www.univearths.fr/en} project
      ``Impact of black holes on their environment''. 
        Partial support through the COST action MP0905 Black Holes in a Violent Universe is acknowledged.
\end{acknowledgements}

\bibliographystyle{aa}
\bibliography{biblio}

\begin{appendix}
\section{regions}
\subsection{Regions excluded from analysis}
To avoid pollution by bright point sources, we have excluded circular regions around a number of bright objects.
Transients objects are removed only during periods when their emission is bright. The list of removed sources is given
in Table \ref{tab:trans}. 

\begin{table}
  \caption{Exclusion regions corresponding to known bright transient and persistent sources in the CMZ. }
\centering
\begin{tabular}{l c c c }
\hline\hline
Source & R.A. (\degr) & Dec. (\degr) & Radius ($''$)\\
\hline
 AX~J1745.6--2901 & 359.9220 & --0.0400 & 15\\  
 GRS~1741.9--2853 & 359.9526313 &  0.1202 & 90 \\       
 GRO~1744-28 &  0.0443 &  0.3014 & 40\\                 %
 XMMU~J174445.5--295044 & 359.1280 & --0.3144 & 100 \\ %
 XMMU~J174554.4--285456 & 0.0503 & --0.0428 & 60\\ 
 XMMU~J174654.1--291542 & 359.8672 & --0.4091 & 30 \\ %
 1E~1740.7--2942 & 359.1158 & --0.1057 & 210\\  %
 1E~1743.1--2843 & 0.2608 & --0.0287 & 174\\    
 1A~1743--288 & 359.5585 & --0.3880 & 252\\             
 SAX~J1747.0--2853 & 0.2067 & --0.2385 & 150\\  
 IGR~J17497--2821 & 0.9533 & --0.4528 & 360 \\  
 XMMU~J174505.3--291445 & 359.6751 &  --0.0623 & 30\\ 
 CXOGC J174537.9--290025 & 359.9412 & --0.0389 & 25\\ 
\hline
\end{tabular}
\label{tab:trans}
\end{table}

\subsection{6.4 keV bright regions}
Light curves have been extracted in various regions to study variability of the \six\ emission. A large
fraction of these regions have been previously studied and the rest was found thanks to the systematic search for
variability described above. The regions are ellipses and we give their parameters (centre coordinates, radii and
rotation angle) in Table \ref{tab:trans}. We note that the region G0.24--0.17 is a polygon in our analysis. For simplicity
we give the parameters of an ellipse matching as closely as possible its morphology. Regions used to estimate the background
level in different regions of the CMZ are also given.

\begin{table*}
  \caption{Parameters of the regions of interest considered in this work. Regions are grouped by sub-regions in the CMZ.
    The regions used to determine the background emission level are given at the bottom of the table.
    References of some previous studies of these regions are also given. We note that the parameters that were used
    in those studies might not be strictly equal to ours. References:
    (1) \citet{Koyama96},
    (2) \citet{2000ApJ...534..283M},
    (3) \citet{2001ApJ...558..687M},
    (4) \citet{murakami01},
    (5) \citet{2002ApJ...568L.121Y},
    (6) \citet{2006JPhCS..54..133S},
    (7) \citet{2007PASJ...59S.221K},
    (8) \citet{2008PASJ...60S.191N},
    (9) \citet{2009PASJ...61..593F},  
    (10) \citet{nakajima09},
    (11) \citet{inui09},
    (12) \citet{terrier10},
    (13) \citet{ponti10},
    (14) \citet{2010PASJ...62..971N},
    (15) \citet{nobukawa11},
    (16) \citet{capelli11},
    (17) \citet{2012AA...546A..88T},
    (18) \citet{capelli12},
    (19) \citet{clavel13},
    (20) \citet{ryu13},
    (21) \citet{clavel14b},
    (22) \citet{2014ApJ...781..107K},
    (23) \citet{Zhang15}.
    }
\centering
\begin{tabular}{l c c c c c}
\hline\hline
Name & l (\degr) & b (\degr) & Radius ($''$) & angle ($^\circ$)& References\\
\hline
MC1         & 0.021 & -0.052 & 84, 42   & 0    &  13, 18, 19 \\ 
MC2         & 0.031 & -0.077 & 65, 40   & 30   &  13, 18, 19 \\ 
Br1         & 0.064 & -0.076 & 47, 87   & 55   &  13, 19 \\ 
Br2         & 0.106 & -0.083 & 73, 46   & 30   &  13, 19 \\ 
G0.11--0.11 & 0.112 & -0.108 & 97, 176  &  0   &  5, 13, 18, 19 \\ 
Arches      & 0.124 &   0.017 & 25, 59   & 214  &  6, 16, 17, 21, 22\\ 
DX          & 0.111 &   0.075 & 75, 35   & 320  & 16 \\ 
Fil. 2011\footnote{This region has a box shape, not an elliptical one.}  & 0.107  & -0.084 & 26,60 & 0   & 19 \\ 
G0.04--0.13 &  0.043 &  -0.159  &  90    & ---  & 19 \\
G0.02$+$0.01 & 0.017 & 0.01  & 50,140 & 0 & This work \\
G0.00--0.02 & 0.086 & -0.023 & 50,140 & 50 & This work \\
G0.24--0.17 & 0.240 & -0.170 & 490,70   & 0    &   This work \\ 
\hline
\sgrb         & 0.643   & -0.078  & 410       & --- & 1 \\ 
\sgrb2        & 0.665   & -0.027  & 120       & --- & 2, 3, 11, 13, 15, 23 \\ 
G0.74--0.11   & 0.738   & -0.098  & 60,150    & 330 & 7, 11, 15\\ 
G0.66--0.13   & 0.661   & -0.132  & 72,144    & 0   & 23 \\ 
G0.50--0.11   & 0.500   & -0.109  & 80        & --- & 8 \\ 
G0.56--0.11   & 0.565   & -0.117  & 90        & --- & This work\\
\hline
\sgrc         & 359.435 & -0.107  & 194, 380  & 0   & 4\\ 
C1            & 359.440 & -0.065  & 120       & --- & 10, 20 \\ 
C2            & 359.460 & -0.150  & 120       & --- & 10, 20\\ 
C3            & 359.580 & -0.150  & 120, 318  & 0   & 20 \\ 
C4            & 359.424 & -0.157  & 95, 156   & 0   & This work \\
M359.38--0.00 & 359.380 & 0.000    & 90        & --- & 10 \\
M359.22--0.04a& 359.292 & -0.007  & 180, 110  & 320 & 14 \\ 
M359.22--0.04b& 359.228 & -0.071  & 180, 110  & 320 & 14 \\ 
\hline
\sgrd-core    & 1.156   & -0.128  & 213, 188 & 120  & This work\\
\hline
\hline
Sgr A -- bkg  &  359.910 & -0.160  &  285   &      & \\
Sgr B -- bkg  &  0.490   & 0.008   &  280   &      & \\
Sgr C -- bkg  &  359.700 & -0.090  &  360,610 & 0  & \\
Sgr D -- bkg  & 1.065    & -0.015  &  280   &      & \\
M359.22 -- bkg & 359.308 & -0.143  & 240    &      & \\
\hline
\end{tabular}
\label{tab:regions}
\end{table*}

\end{appendix}

\end{document}